\documentclass[sigplan,twocolumn,10pt,nonacm]{acmart}

\acmSubmissionID{343}
\usepackage{xurl}
\usepackage{subfig}
\usepackage{graphicx}
\usepackage[inline]{enumitem}
\setitemize[1]{itemsep=0pt,partopsep=0pt,topsep=5pt,parsep=\parskip}
\usepackage{cancel}

\usepackage{ulem}
\usepackage{xspace}
\usepackage{tikz}
\usepackage{pifont}
\usepackage{filecontents}
\usepackage{tabularx}
\usepackage{booktabs}
\usepackage{todonotes}
\usepackage{multirow}
\usepackage{circledsteps}
\usepackage{algorithm}
\usepackage{xcolor}

\usepackage{algorithmicx}
\usepackage{algpseudocode} 

\usepackage{amsmath}

\usepackage{amssymb}  

\iftrue 

\else

\fi

\newcommand{\prjname}{\textsf{DisagFusion}\xspace}

\definecolor{comment-blue}{rgb}{0,0,1}

\begin{document}
\raggedbottom

\title[\prjname]{\prjname: Asynchronous Pipeline Parallelism and Elastic Scheduling for Disaggregated Diffusion Serving}



\author{Hantian Zha$^{1}$\char44\ Teng Ma$^{2}$\char44\ Yang Yong$^{3}$\char44\ Haiwen Fu$^{3}$\char44\ Ruiyang Ma$^{4}$\char44\ Wei Gao$^{5}$\char44\ Ruihao Gong$^{6}$\char44\ Xianglong Liu$^{6}$\char44\ Wei Wang$^{5}$\char44\ Yunpeng Chai$^{1}$}

\affiliation{%
  \institution{
    $^1$Renmin University of China \quad 
    $^2$Independent Researcher \quad 
    $^3$SenseTime \quad 
    $^4$Peking University \quad \\
    $^5$Hong Kong University of Science and Technology \quad 
    $^6$Beihang University
  }
  \country{}
}





\maketitle

\pagestyle{plain}


\section*{Abstract}
Diffusion-based generation is increasingly powering production content pipelines; however, deploying these models at scale remains a significant challenge. Model weights frequently exceed the memory capacity of commodity GPUs, while the encoder, diffusion transformer (DiT), and decoder stages exhibit highly imbalanced computational and memory footprints. A natural remedy is disaggregated serving---running stages as separate services on heterogeneous GPUs---yet this introduces new bottlenecks, including stage handoff overheads and fast-changing workloads that make cross-stage provisioning and scheduling brittle.

This paper presents \prjname{}, enabling asynchronous pipeline parallelism and elastic scheduling for disaggregated diffusion serving. First, \prjname{} introduces asynchronous pipeline parallelism that overlaps computation and stage-to-stage communication to reduce pipeline bubbles and mitigate network jitter. Second, \prjname{} employs a hybrid instance scheduling strategy that combines lightweight performance prediction with runtime feedback to continuously rebalance instance ratio across stages under workload shifts.
We implement \prjname{} and evaluate it with modern diffusion models. Compared to a monolithic baseline, \prjname{} improves throughput by 3.4$\times$--20.5$\times$ and reduces end-to-end latency by 18.5$\times$, while enabling flexible, cost-efficient deployment across heterogeneous GPUs.

\section{Introduction}

Generative AI has rapidly evolved from research prototypes into production workflows spanning chatbots and coding assistants~\cite{brown2020gpt3}, and increasingly multimodal creation such as text-to-image~\cite{saharia2022imagen,rombach2022ldm} and text-to-video generation~\cite{ho2022imagenvideo,singer2022makeavideo,villegas2022phenaki}. Within this landscape, diffusion-based models—often built upon denoising backbones and diffusion transformers—have emerged as the dominant paradigm for high-fidelity synthesis~\cite{ho2020ddpm,peebles2023dit}. These models inherently exhibit a decomposable execution structure comprising three distinct stages: an \textbf{E}ncoder stage, a denoising Diffusion \textbf{T}ransformer, and a \textbf{D}ecoder stage~\cite{ho2020ddpm,rombach2022ldm,peebles2023dit,ho2022videodiffusion,ho2022imagenvideo,singer2022makeavideo,villegas2022phenaki,blattmann2023videoldm,stability2023stablevideodiffusion}. Each stage needs to load model weights and activations, consuming substantial GPU memory. Moreover, in many production settings, generation is performed in an offline manner and the final output is returned only after completion, making the workload less sensitive to per-step latency than interactive inference~\cite{lin2025understanding}.

However, running such models on a single machine faces several practical problems:
\textit{(1) Models with a large number of parameters often cannot fit entirely in GPUs with limited memory capacity.} Most models have large parameter sizes and cannot be fully loaded on consumer-grade GPUs with 24~GB VRAM, and some even exceed the memory capacity of professional GPUs with 80~GB VRAM. As a result, inference often has to dynamically load the required weights on demand, which can severely degrade system performance.
\textit{(2) Distinct pipeline stages exhibit heterogeneous runtime characteristics.} Encoder and decoder stages exhibit lower arithmetic intensity, making them suitable for consumer-grade GPUs, whereas DiT stages tend to be compute-bound and benefit from the higher throughput of professional GPUs~\cite{peebles2023dit}. If all stages are executed on the same machine, the GPU's compute resources cannot be fully utilized.

To address the above issues, we adopt a decoupled (disaggregated) architecture that splits the encoder, DiT, and decoder into independent services and deploys them on different GPUs. This disaggregated deployment brings three key benefits:
\begin{itemize}
  \item It can remove cross-stage interference and improve throughput. By isolating the encoder, DiT, and decoder onto separate GPUs, each stage can run in its own kernel, memory pool, and concurrency control.
  \item It increases resource utilization and reduces hardware cost. Heterogeneous stages can be mapped to heterogeneous GPUs (e.g., compute-dense DiT on high-end GPUs and memory, while lightweight stages on commodity GPUs) to better match price--performance.
  \item It enables flexible instance provisioning across stages. Because the three stages are decoupled, we can independently scale the number of instances for the encoder, DiT, and decoder to better balance the pipeline and achieve higher throughput.
\end{itemize}

However, after adopting disaggregated deployment, we must also address several new challenges.

\textit{\textbf{1) Disaggregated serving requires careful workflow design to avoid the network becoming the bottleneck.}} With stage-to-stage communication on the critical path, network jitter directly translates into longer handoff time. Moreover, since the encoder and decoder are relatively short, they are typically provisioned with far fewer instances than the DiT, and a synchronous send/receive pattern can further amplify tail latency and reduce overall throughput~\cite{narayanan2020clockwork,crankshaw2020inferline}.

\textit{\textbf{2) In generation services, the number of requests and request parameters vary over time.}} For instance, the request volume can exhibit a 2.2$\times$ peak-to-valley gap~\cite{coppock2025lithos, lin2025understanding, qiu2025modserve}, and request parameters (e.g., output resolution, and denoise steps) also change dynamically~\cite{luodit,soprompts_sora_runway_pika,vidwave_pika_vs_sdv}. On an A10, running 50-step inference of the Wan2.2 model takes 930~s, while 4-step distilled inference takes only 74.1~s. In contrast, the encoder and decoder are relatively stable (e.g., 5.46~s and 9.62~s). This indicates that the optimal instance ratio across stages can differ significantly under different workloads.

Therefore, we present \prjname{} (code: \url{https://github.com/ModelTC/LightX2V}), enabling asynchronous pipeline parallelism and elastic scheduling for disaggregated diffusion serving. \prjname{} \ding{182} devises an asynchronous pipeline parallelism mechanism that seamlessly overlaps computation with inter-stage communication, eliminating pipeline bubbles induced by communication overheads; and \ding{183} introduces a hybrid instance scheduling strategy that synthesizes static performance modeling with dynamic runtime feedback to achieve elasticity, sustaining near-optimal throughput even amidst volatile workload shifts.

In summary, we make the following contributions:
\begin{itemize}
  \item We analyze the benefits of disaggregated architecture and then systematically study two key challenges for disaggregated diffusion generation serving (\S~\ref{sec:bg_motivation}).
  \item To address these challenges, we propose \prjname{} with two techniques: asynchronous pipeline parallelism for stage-wise computation overlap, and hybrid-strategy instance scheduling for heterogeneous workload (\S~\ref{sec:design}).
  \item We implement \prjname{} and compare it against the monolithic baseline. \prjname{} reduces end-to-end latency by 18.5$\times$, and improves throughput by 3.4$\times$--20.5$\times$ over the baseline (\S~\ref{sec:evaluations}).
\end{itemize}

\section{Background and Motivation}
\label{sec:bg_motivation}

We outline the diffusion architecture and its workload characteristics (\S\ref{sec:bg-arch}, \S\ref{sec:bg-workload}), then contrast monolithic and disaggregated deployments to identify key challenges and opportunities (\S\ref{sec:bg-mono-disagg}, \S\ref{sec:bg-chal-oppo}). Finally, we discuss the limitations of existing serving techniques for this domain (\S\ref{sec:bg-ineff}).

\subsection{Diffusion Model Architecture}\label{sec:bg-arch}
\begin{figure}[t]
  \centering
  \includegraphics[width=\linewidth]{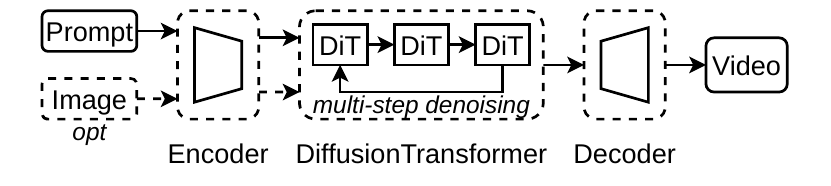}
  \caption{Video Generation Pipeline based on Diffusion Transformers.}
  \label{fig:diffusion_arch}
\end{figure}

As illustrated in Figure~\ref{fig:diffusion_arch}, generation models typically follow an Encoder-Transformer-Decoder architecture~\cite{rombach2022ldm,peebles2023dit}. The process begins with an Encoder stage that projects input conditions, such as text prompts and optional images, into a latent space. The core generation occurs within the Diffusion Transformer (DiT), which iteratively performs denoising on the latent features to synthesize temporal dynamics. Finally, a Decoder reconstructs the high-fidelity output frames from the refined latents. Among these stages, the DiT module serves as the computational bottleneck, requiring extensive iterative processing to ensure output quality.

\subsection{Workload Characteristics}\label{sec:bg-workload}
In this subsection, we analyze the model weights and compute intensity across different pipeline stages. Our measurements are collected on machines with different hardware configurations, and all runs use 50-step inference to ensure a consistent comparison.

\begin{figure}[t]
  \centering
  \includegraphics[width=\linewidth]{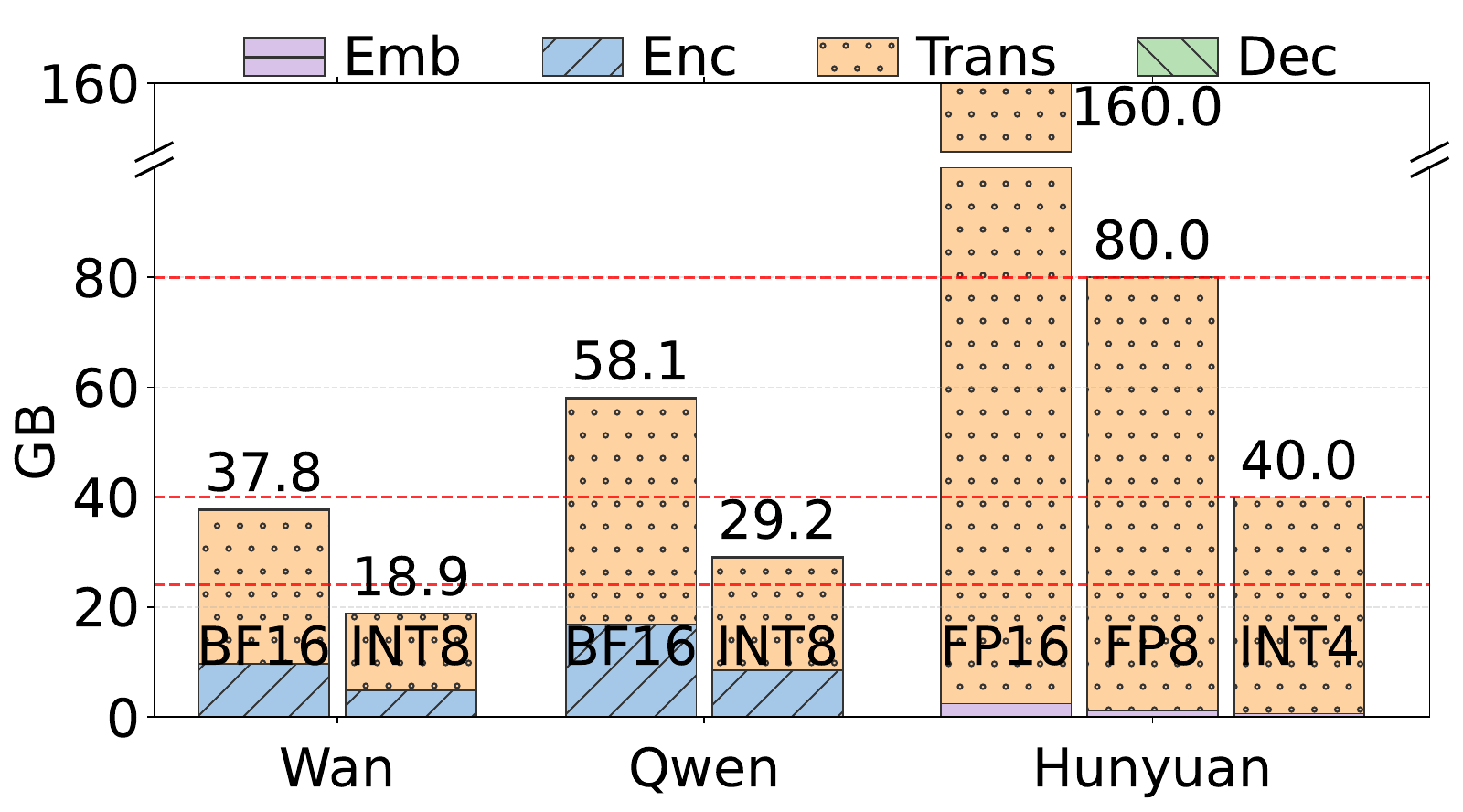}
  \caption{Memory footprint of different models. (The red lines indicate GPU memory capacities: 24~GB for RTX~4090 and A10, 40~GB for A100, and 80~GB for A100 and H100.)}
  \label{fig:model-memory-bg}
\end{figure}

\textbf{Substantial model weight memory consumption.} Models with a large number of parameters often cannot fit entirely in GPUs with limited memory capacity. As shown in Figure~\ref{fig:model-memory-bg}, Wan is used for both T2V and I2V inference, whereas Qwen and Hunyuan are used for T2I inference. For the Wan~2.1 14B model~\cite{wan21-t2v-14b}, BF16 weights require 37.8~GB of GPU memory, which already cannot fit on consumer-grade GPUs such as NVIDIA A10 and RTX~4090; for the Qwen~2512 model~\cite{qwen-image-2512}, the footprint reaches 58.1~GB, which cannot fit even on a 40~GB A100; even after quantization, the model remains too large to reside entirely on a single consumer-grade GPU (e.g., NVIDIA A10 or RTX 4090 with 24 GB VRAM). If using the Hunyuan model~\cite{tencent-hunyuan-video}, even FP16 and FP8 precision cannot fit in the memory of any currently available GPU. When GPU memory is insufficient, parts of the model must be offloaded to host memory during inference, incurring non-negligible CPU--GPU data transfer overhead~\cite{sheng2023flexgen}, and multi-GPU model parallelism introduces substantial inter-GPU communication overhead that limits scalability~\cite{li2023alpaserve,yao2022deepspeedinference}.

\textbf{Heterogeneous compute intensity across stages.} The computational complexity of generation is predominantly governed by the DiT. The DiT \ding{182} has a much larger parameter size (e.g., Wan~2.1-14B in BF16 has 28.0~GB of DiT weights), and \ding{183} its attention computation scales as $O(T^2\cdot D)$ for tokens $T$ and hidden dimension $D$. In addition, \ding{184} the denoising process requires multi-step iterations---typically 40--50 steps for standard samplers, and even distilled models such as LightX2V still require 4 or 8 steps~\cite{lightx2v}. In contrast, the encoder and decoder use only 9.6~GB and 0.1~GB of weights, respectively. Their computation scales roughly linearly with the number of pixels $I$ (approximately $O(I^2)$ in practice), and they do not require iterative execution. Together, these factors make the DiT dominate the end-to-end runtime.

\subsection{Monolithic vs. Disaggregated}\label{sec:bg-mono-disagg}
\begin{figure}[t]
  \centering
  \includegraphics[width=\linewidth]{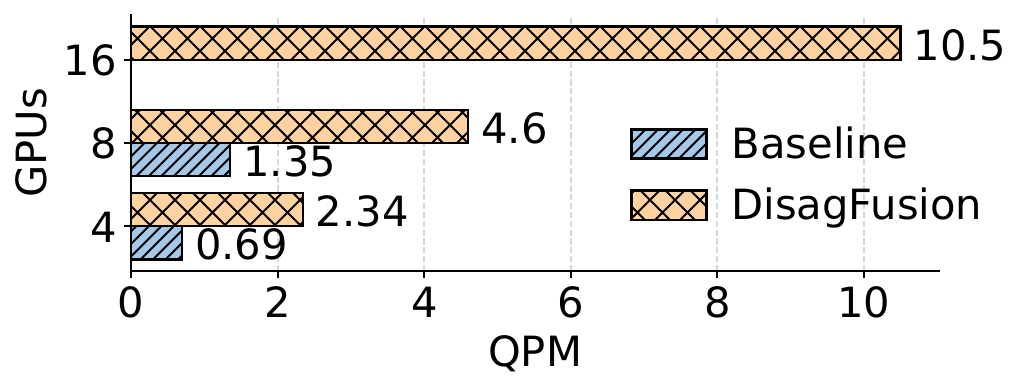}
  \caption{Comparison of scalability for different deployment strategies (baseline and the disaggregated version).}
  \label{fig:scalability_wan_A10_4step}
\end{figure}

As shown in Figure~\ref{fig:scalability_wan_A10_4step}, we evaluate a monolithic LightX2V deployment as the baseline and compare it against the disaggregated version. The result shows that the disaggregated version achieves higher throughput and better scalability. In the 8-GPU setting, its throughput is 3.41$\times$ of the baseline. This is because the weights of the three stages cannot fit entirely in GPU memory, so the monolithic deployment must load models for each stage onto the GPU when it runs, incurring substantial I/O overhead. In contrast, the disaggregated deployment loads each stage's weights once onto its dedicated GPU and keeps them resident, thereby avoiding repeated loading/unloading, better utilizing GPU resources.

\begin{figure}[t]
  \centering
  \includegraphics[width=\linewidth]{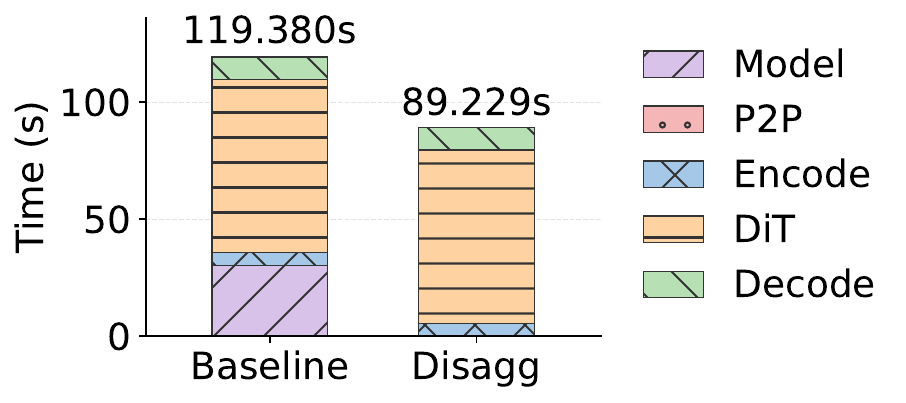}
  \caption{Single-request end-to-end latency breakdown of the baseline and the disaggregated version (4-step inference). (\textit{Model} denotes model loading/unloading overhead; \textit{Encode}, \textit{DiT}, and \textit{Decode} denote computation time of three stages; \textit{P2P} denotes inter-stage data transfer overhead.)}
  \label{fig:latency_breakdown}
\end{figure}

To analyze the system performance bottlenecks, we measure the end-to-end latency breakdown for a single 4-step inference request under both the monolithic baseline and the disaggregated version (Figure~\ref{fig:latency_breakdown}). The baseline spends an extra 30.3\,s on model loading/unloading, accounting for 25.3\% of the total latency; the disaggregated version avoids this overhead by keeping each stage's model resident and thus eliminates repeated loading. Under a stable network, the end-to-end latency of the disaggregated version is dominated by computation in the encoder, transformer, and decoder stages, with the transformer accounting for most of the time (83\%). This motivates a disaggregated deployment for diffusion model serving.

\begin{figure}[t]
  \centering
  \includegraphics[width=\linewidth]{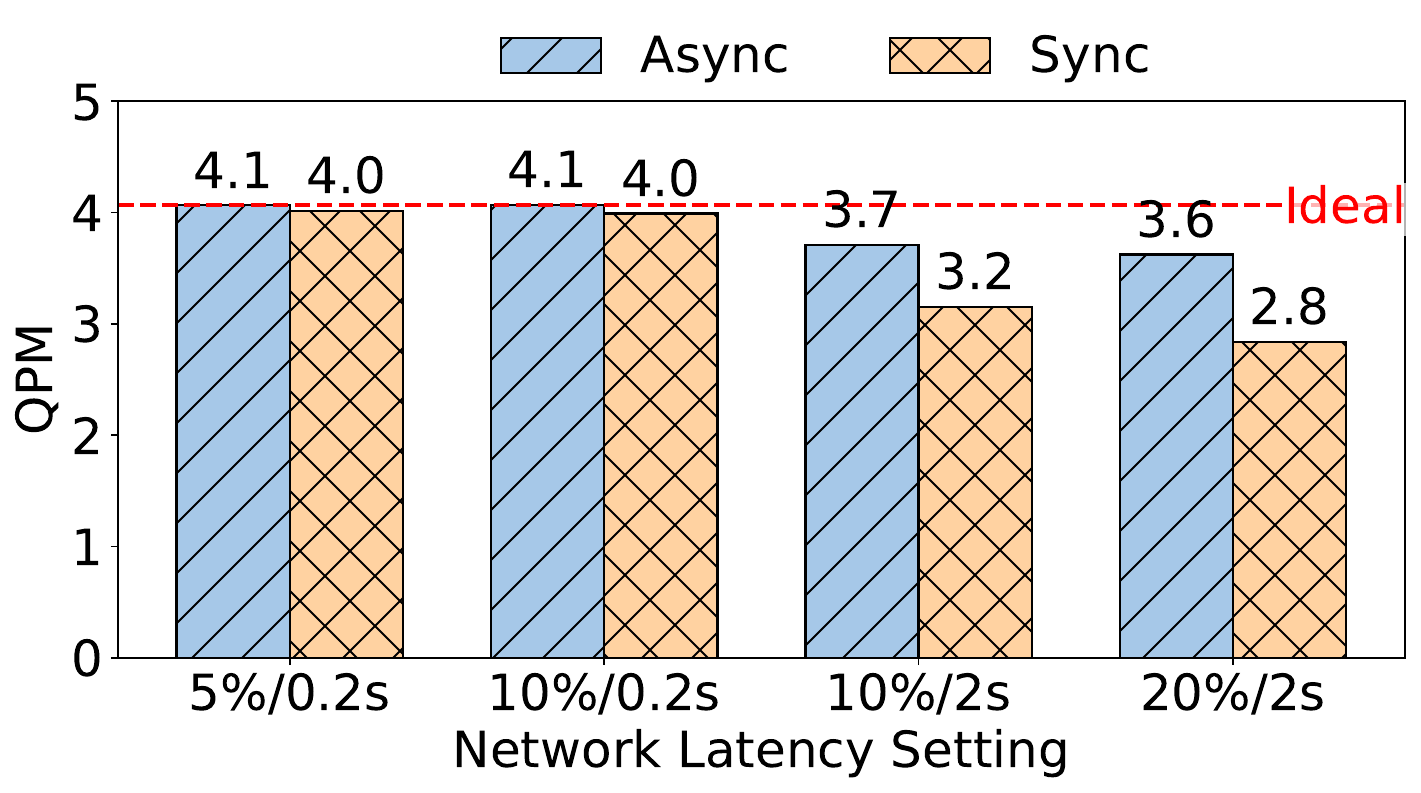}
  \caption{Impact of network latency/jitter on synchronous inter-stage transfer. Here, ``5\%/0.2s'' means that each transfer via the transfer engine has a 5\% probability of incurring an additional 0.2-second delay.}
  \label{fig:network_latency}
\end{figure}

\subsection{Challenges and Opportunity}\label{sec:bg-chal-oppo}
By adopting a disaggregated deployment strategy, our design effectively circumvents the model loading bottleneck while incurring negligible overhead. However, this architectural shift introduces two fundamental challenges.

\paragraph{Challenge 1: Disaggregated serving requires careful workflow design to avoid the network becoming the bottleneck.}
A disaggregated architecture splits the model into independent services, incurring cross-node inter-stage communication. However, traditional synchronous coordination severely degrades performance and scalability.
When stages pass intermediate results synchronously (i.e., the upstream stage blocks until the downstream stage receives them), transient network jitter can directly translate into bubble time and backpressure across the whole pipeline, significantly hurting tail latency and throughput (Figure~\ref{fig:network_latency}). Even under mild fluctuations (10\%/2s), throughput drops by 22.5\%; under severe fluctuations (20\%/2s), throughput drops by 30.3\%.

\paragraph{Opportunity 1: Asynchronous communication can eliminate pipeline bubbles caused by network latency.}
Synchronous handoffs force the sender to stall until the downstream stage acknowledges receipt, turning every network fluctuation into GPU idle time. In contrast, asynchronous communication decouples producer and consumer: the sender proceeds to the next request immediately after dispatching the tensor, overlapping computation with in-flight transfers. Transient jitter is therefore absorbed by inter-stage buffering rather than amplified into pipeline bubbles. The benefit is substantial—even under severe network conditions, throughput falls by only 11.0\%, far less than the 30.3\% drop suffered by the synchronous baseline.

\begin{table}[t]
  \caption{Execution time of the config across stages. Other parameters use the default settings in the LightX2V project.}
  \label{tab:stage_time}
  \centering
  \small
  \begin{tabular}{@{}lcccc@{}}
    \toprule
    \textbf{Steps} & \textbf{Resolution} & \textbf{Enc (s)} & \textbf{DiT (s)} & \textbf{Dec (s)} \\
    \midrule
    50-steps & 832$\times$480 & 5.46 & 930  & 9.62 \\
    8-steps  & 832$\times$480 & 5.46 & 149  & 9.62 \\
    4-steps  & 832$\times$480 & 5.46 & 74.1 & 9.62 \\
    1-step   & 832$\times$480 & 5.46 & 18.7 & 9.62 \\
    \bottomrule
  \end{tabular}
\end{table}

\paragraph{Challenge 2: In generation services, the number of requests and request parameters vary over time.}
Table~\ref{tab:stage_time} reports the per-stage latency of the Wan2.2 model under different step counts. Under 4-step distillation, the encoder, DiT, and decoder stages take approximately 5 s, 75 s, and 10 s, respectively; under 1-step distillation, the corresponding latencies shift to 5 s, 20 s, and 10 s. Because the DiT stage dominates in the first scenario while becoming far less constraining in the second, the bottleneck stage moves across configurations. Consequently, no fixed encoder-to-DiT-to-decoder instance ratio can sustain peak throughput under both workloads.

\begin{figure}[t]
  \centering
  \includegraphics[width=\linewidth]{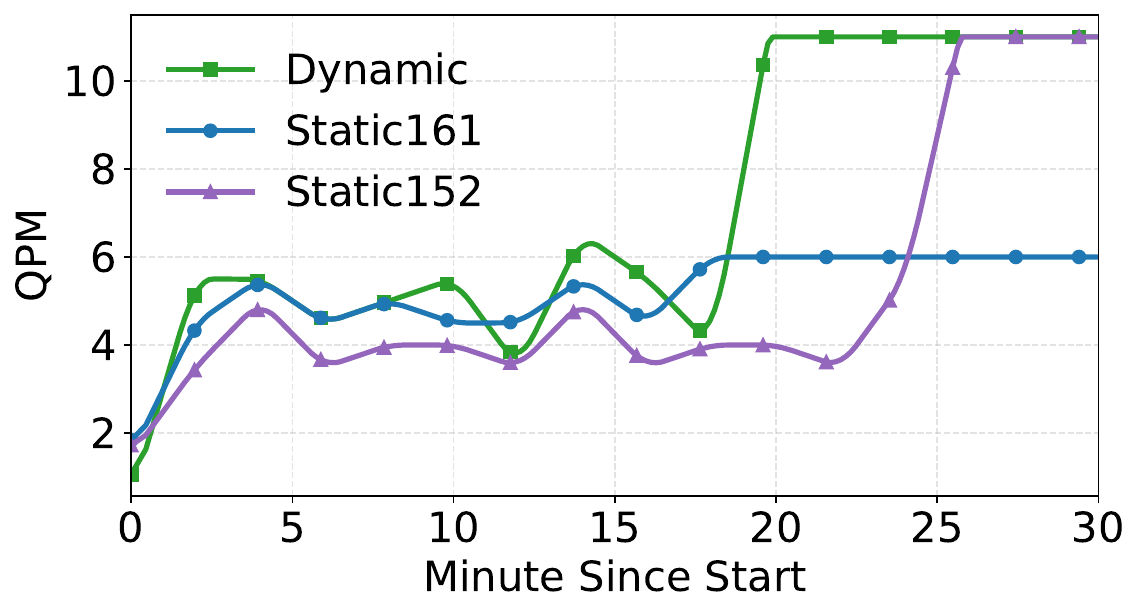}
  \caption{Real-time throughput under varying request parameters. The first 15 minutes use 4-step distill requests, after 15 minutes, the requests switch to 1-step distill. Static161 denotes a static 1:6:1 instance ratio, Static152 is defined similarly, and Dynamic denotes dynamic instance scheduling.}
  \label{fig:realtime_thrput_p}
\end{figure}

\paragraph{Opportunity 2: Dynamic instance scheduling can maximize throughput under varying workloads.}
To highlight the time-varying nature of real-time serving workloads, we evaluate different workloads on an 8-GPU testbed. Figure~\ref{fig:realtime_thrput_p} shows the measured throughput under changing request parameters. Specifically, we send 4-step requests in the first 15 minutes, and switch to 1-step requests after 15 minutes. For 4-step requests, the 1/6/1 instance configuration achieves the highest throughput, whereas for 1-step requests, the 1/5/2 configuration achieves the highest throughput. Therefore, by dynamically adjusting instances to use the 1/6/1 configuration in the first 15 minutes and the 1/5/2 configuration after 15 minutes, we can achieve the maximum throughput under both workloads.

\subsection{Inefficiencies of Existing Works}\label{sec:bg-ineff}

Disaggregated serving has been extensively explored in the context of LLMs, yielding techniques like KV-cache-centric memory management and prefill--decode splitting~\cite{vllm2023pagedattention,qin2024mooncake,yu2022orca,zhong2024distserve}. However, these optimizations rely on assumptions that do not hold for diffusion generation. We analyze their limitations from two perspectives:

\textbf{Inter-stage Latent Communication.} In disaggregated LLM serving, cross-instance transfer involves small, regular KV cache blocks, making transfer latency negligible and synchronous handoffs tolerable~\cite{vllm2023pagedattention,zhong2024distserve,qin2024mooncake}. In contrast, diffusion serving requires transmitting full intermediate latent tensors between the encoder, DiT, and decoder. These tensors are orders of magnitude larger and scale with output resolution and frame count. Synchronous transfers in this context cause upstream stages to block on network jitter, stalling the entire pipeline. Therefore, an asynchronous execution model overlapping communication with computation is essential to prevent throughput degradation.

\textbf{Dynamic Stage Bottlenecks.} Disaggregated LLM systems typically rely on static provisioning, as the prefill and decode phases have relatively stable characteristics, and decode iteration counts are inherently unpredictable~\cite{yu2022orca,narayanan2020clockwork,zhong2024distserve,qin2024mooncake}. Conversely, diffusion generation offers deterministic per-stage execution times: the encoder and decoder run once with stable latency, and the DiT stage executes a user-specified number of denoising steps. Static strategies borrowed from LLMs fail to exploit this predictability, as the optimal instance ratio shifts continuously with request parameters (e.g., step count, resolution). Consequently, a hybrid instance scheduling mechanism---combining static performance prediction with dynamic runtime feedback---is required to continuously rebalance the number of stage instances against evolving workloads.


\section{System Design}\label{sec:design}

This paper presents \prjname{}, a scalable and efficient serving system designed for diffusion models in disaggregated architectures. \prjname{} co-designs pipeline execution and resource management to actively combat the two dominant overheads of disaggregation: (1) synchronous inter-stage communication stalls GPU execution under network jitter, and (2) static instance allocations cannot adapt to dynamic workloads.

We explain our asynchronous pipeline parallelism and hybrid-strategy instance scheduling in Sections~\ref{sec:decentral-pipeline} and~\ref{sec:hybrid-scheduling}.

\begin{itemize}
  \item \textbf{Asynchronous pipeline parallelism (\S~\ref{sec:decentral-pipeline}).} With the co-design of decentralized inter-stage coordination and asynchronous pipeline execution, we eliminate centralized bottlenecks and enable computation-communication overlap.
  \item \textbf{Hybrid-strategy instance scheduling (\S~\ref{sec:hybrid-scheduling}).} Our monitoring framework continuously aggregates service metrics and guides elastic scheduling decisions through a policy that fuses static performance models with dynamic runtime observations.
\end{itemize}

\subsection{Overview}
\begin{figure}[t]
  \centering
  \includegraphics[width=\linewidth]{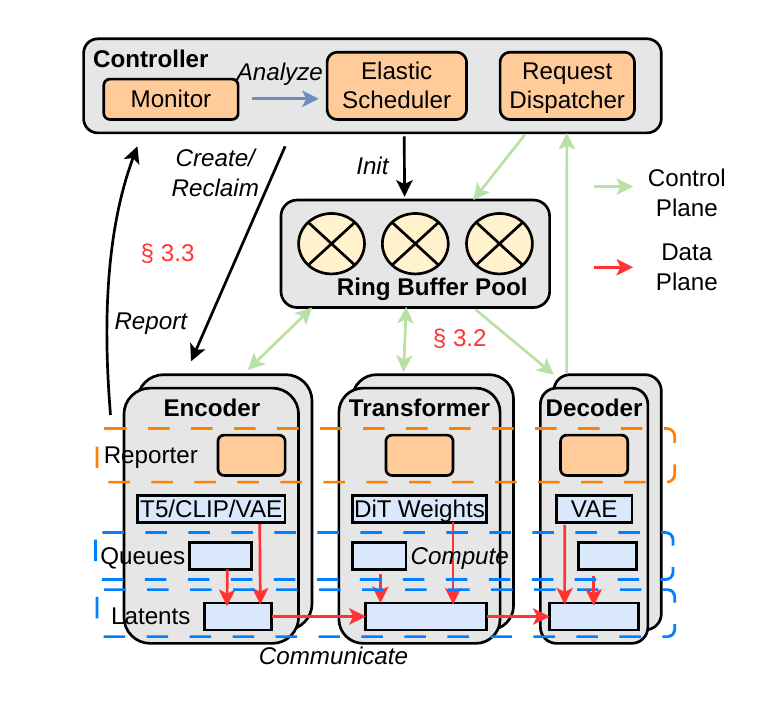}
  \caption{The overall architecture of \prjname{}.}
  \label{fig:architecture}
\end{figure}

Figure~\ref{fig:architecture} presents the overall system architecture. System architecture comprises two main optimization modules: Asynchronous Pipeline Parallelism for request processing workflow and Hybrid-strategy Instance Scheduling for dynamic resource management.

\subsubsection{Asynchronous Pipeline Parallelism}
The \textit{control plane} is centered around the Ring Buffer Pool, which manages request metadata and service metadata. Services communicate through a producer-consumer model where each stage acts as both producer and consumer. The Controller is responsible for initial request dispatching, while other services alternately transmit metadata to complete computation in their respective stages.

The \textit{data plane} consists of three computation stages (Encode, Diffusion Transformer, and Decode) that process data through a decentralized pipeline. These stages exchange intermediate tensors via the mooncake transfer engine, enabling zero-copy tensor transmission. The Encoder produces latent tensors, the Transformer consumes and refines them through iterative denoising, and the Decoder consumes the refined tensors to reconstruct the final output. This producer-consumer architecture supports asynchronous execution with computation-communication overlap.

\subsubsection{Hybrid-strategy Instance Scheduling}
A dynamic scheduling mechanism combines static performance predictions with dynamic runtime indicators to enable adaptive resource provisioning. The Controller continuously monitors system metrics and makes scheduling decisions. For instance, scaling and resource allocation across stages based on workload characteristics and performance requirements. This hybrid approach achieves scalability, efficiency, and adaptability in disaggregated environments.

\subsection{Asynchronous Pipeline Parallelism}\label{sec:decentral-pipeline}

\begin{figure}[t]
  \centering
  \includegraphics[width=\linewidth]{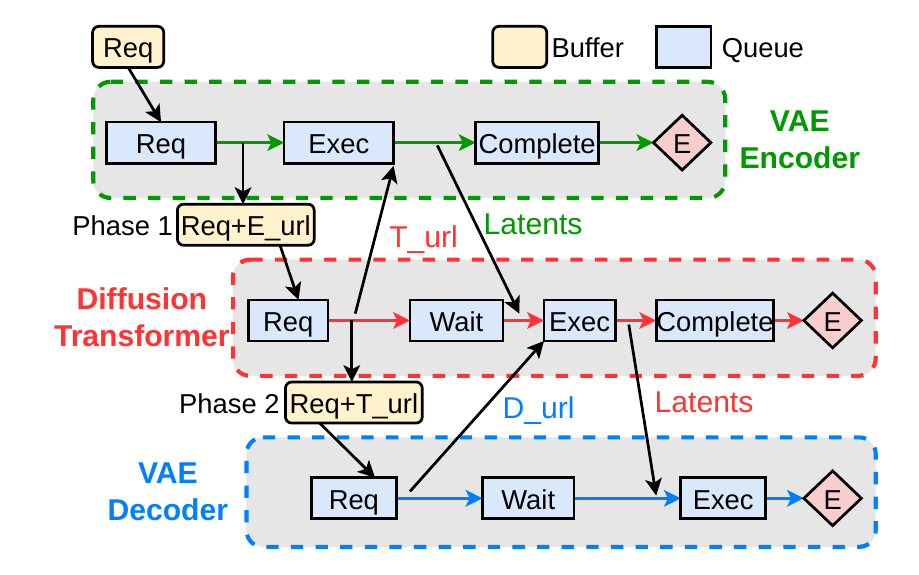}
  \caption{A request workflow between three stages.}
  \label{fig:pipeline-parallelism}
\end{figure}

As shown in Figure~\ref{fig:pipeline-parallelism}, \prjname{} organizes each generation request into an pipeline that mirrors a diffusion model’s computation graph: (1) Encoder, which preprocesses conditioning inputs (e.g., text prompts, negative prompts, and control signals) and produces conditioning hidden states (e.g., T5, CLIP, VAE encoder outputs); (2) Transformer, which runs the main denoising backbone (e.g., a Transformer network) for iterative diffusion timesteps and updates the latent representation; (3) Decoder, which decodes the final latent into RGB output frames (e.g., via a VAE decoder) and performs lightweight post-processing before returning the output.

\noindent\textbf{Encoder stage (request admission and conditioning preparation).}
Upon arrival, the request scheduler inserts the request into the global \textit{request buffer}. An encoder instance dequeues the request into its local \textit{request queue}, and a worker thread then initializes the request (on-demand model loading and GPU memory allocation). Next, the encoder produces the request metadata into the \textit{phase1 buffer} and enqueues the request into the \textit{execute queue}. The \textbf{Transformer} instance that fetches the corresponding metadata asynchronously sends a destination address to the encoder; the encoder worker proceeds to compute and, upon completion, sends intermediate results to the \textbf{Transformer} and moves the request into the \textit{complete queue}. After the send is acknowledged, the encoder dequeues it from \textit{complete queue} and releases resources.

\noindent\textbf{Transformer stage (iterative denoising and cross-stage interaction).} A transformer instance dequeues the request metadata, enqueues the request into its local \textit{request queue}, and a worker thread initializes it (on-demand model loading, GPU memory allocation, and sending its address to the \textbf{Encoder}). The transformer then places metadata into the \textit{phase2 buffer} and enqueues the request into the \textit{waiting queue} to await \textbf{Encoder} outputs. The \textbf{Decoder} instance fetches the corresponding metadata asynchronously and sends a destination address to the transformer. Once \textbf{Encoder} outputs are received, the transformer moves the request to the \textit{execute queue} for computation; after finishing, it sends intermediate results to the \textbf{Decoder}, transitions the request into the \textit{complete queue}, and finally dequeues it after successful sending to free resources.

\noindent\textbf{Decoder stage (latent decoding, post-processing, and response).}
A decoder instance dequeues the request metadata into its local \textit{request queue}; a worker thread initializes it (on-demand model loading, GPU memory allocation, and sending its address to the \textbf{Transformer}). The request is then put into the \textit{waiting queue} for \textbf{Transformer} outputs. After receiving the results, the decoder moves it into the \textit{execute queue} to decode the latent into output frames. Once the final output is returned to the request scheduler, the decoder releases resources.

The three stages communicate via asynchronous queues, allowing different requests to occupy different stages concurrently. As a result, each request can flow through the pipeline without being blocked by others at the same stage, and the overall processing becomes fully overlapped.

\subsection{Hybrid-Strategy Instance Scheduling}\label{sec:hybrid-scheduling}

When deploying the service in a cluster, we need to allocate a certain number of instances to the three stages. For simplicity, we assume each instance exclusively occupies one GPU. Table~\ref{tab:hybrid-notation} summarizes the key variables. Let $g_E$, $g_T$, and $g_D$ denote the numbers of GPUs allocated to the Encoder, Transformer, and Decoder stages, respectively. Then we have the resource constraint:

\begin{equation}
  g_E + g_T + g_D \le G.
\end{equation}

For each GPU, we also require that the model and activation memory footprints fit in GPU memory:

\begin{equation}
  S_M + S_A < C.
\end{equation}

For a single request, the execution time of each stage can be modeled as the sum of computation time and communication time:

\begin{align}
  T_E &= \frac{S_{A,E} \cdot I_E}{P_E} + \frac{S_{A,E}}{B_E}, \\
  T_T &= \frac{S_{A,T} \cdot I_T}{P_T} + \frac{S_{A,T1}}{B_{T1}} + \frac{S_{A,T2}}{B_{T2}}, \\
  T_D &= \frac{S_{A,D} \cdot I_D}{P_D} + \frac{S_{A,D}}{B_D}.
\end{align}

Given the per-stage service rates, end-to-end throughput is determined by the minimum rate across all stages:

\begin{equation}
  \mathrm{QPS} = \min\left\{\frac{g_E}{T_E}, \frac{g_T}{T_T}, \frac{g_D}{T_D}\right\}.
\end{equation}

\begin{table}[t]
  \caption{Notations used in the instance scheduling model.}
  \label{tab:hybrid-notation}
  \centering
  \small
  \begin{tabular}{@{}lll@{}}
    \toprule
    \textbf{Variable} & \textbf{Meaning} & \textbf{Unit} \\
    \midrule
    $G$ & Total number of GPUs & -- \\
    $C$ & GPU memory capacity & GB \\
    $S_M$ & Model memory footprint & GB \\
    $S_A$ & Activation memory footprint & GB \\
    $I$ & Compute intensity & FLOPs/byte \\
    $P$ & GPU performance & TFLOP/s \\
    $B$ & Communication bandwidth & GB/s \\
    $g$ & Number of GPUs used & -- \\
    $T$ & Execution time & s \\
    $\mathrm{QPS}$ & Throughput & requests/s \\
    \bottomrule
  \end{tabular}
\end{table}

Therefore, to maximize throughput, we should avoid the bottleneck effect across stages. Optimal instance allocation should approximately balance the per-stage service rates:

\begin{equation}
  \frac{g_E}{T_E} \approx \frac{g_T}{T_T} \approx \frac{g_D}{T_D}.
\end{equation}

For a static workload, $T_E$, $T_T$, and $T_D$ can be pre-computed and are thus easier to predict, making the required ETD instance ratio largely stable. However, under high-concurrency scenarios, request parameters exhibit heterogeneity, request arrival rates vary temporally, and the availability of heterogeneous GPUs changes dynamically.
We monitor GPU utilization, queue length, and queueing delay. When service exhibits high GPU utilization coupled with a sustained growth in both queue length and queueing delay, we trigger a scale-out operation by provisioning additional instances to alleviate the load. Conversely, when an instance demonstrates low GPU utilization and maintains an empty queue over a monitoring period, we initiate a scale-in operation by de-provisioning that instance.

Since these metrics are inherently reactive and may fail to capture abrupt workload spikes, we complement the feedback loop with a predictive layer. By learning the mapping between historical workload characteristics (e.g., request parameters) and the optimal service ratio, the system can proactively reconfigure resources in anticipation of demand shifts. In response to significant workload fluctuations over time, we leverage the trained model to forecast the required instance count and proactively adjust system capacity.

\begin{algorithm}[t]
\caption{Hybrid-strategy instance scheduling}
\label{alg:hybrid-scheduling}
\begin{algorithmic}[1]
\Require Monitoring interval $\Delta$; scale-out thresholds $U_{\text{high}}, Q_{\text{high}}$; scale-in thresholds $U_{\text{low}}$; a workload-change detector $\textsc{Changed}(\cdot)$; a learned predictor $\hat{g}(\cdot)$ mapping workloads to desired instance counts.
\State Initialize instance counts $n_E, n_T, n_D$
\State Initialize history buffer $H \gets \emptyset$
\While{system is running}
  \State Collect metrics $m \gets \{u_s, q_s, d_s\}$ for each service $s \in \{E,T,D\}$
  \State Append $m$ and recent request parameters into $H$

  \If{$\textsc{Changed}(H)$}
    \State Workload features $x \gets \textsc{Featurize}(H)$
    \State $(\hat{n}_E, \hat{n}_T, \hat{n}_D) \gets \hat{g}(x)$
    \State \textsc{Apply}$(\{n_E,n_T,n_D\}, \{\hat{n}_E,\hat{n}_T,\hat{n}_D\})$
    \State \textbf{continue}
  \EndIf

  \ForAll{service $s \in \{E,T,D\}$}
    \If{$u_s > U_{\text{high}}$ \textbf{and} $q_s > Q_{\text{high}}$ \textbf{and} $d_s > d_s'$}
      \State \textsc{ScaleOut}$(s)$
    \ElsIf{$u_s < U_{\text{low}}$ \textbf{and} $q_s = 0$}
      \State \textsc{ScaleIn}$(s)$
    \EndIf
  \EndFor

  \State Sleep for $\Delta$
\EndWhile
\end{algorithmic}
\end{algorithm}

Algorithm~\ref{alg:hybrid-scheduling} summarizes our hybrid scheduling workflow. Here, the monitoring interval $\Delta$ is set to 2\,s by default to accommodate the response time required for cold starts and reclamation. $U_{\text{high}}$ is the scale-out utilization threshold (default 80\%), and $Q_{\text{high}}$ is the scale-out threshold for queue length (default 5), used to scale out when the current GPU utilization exceeds the configured threshold and queued requests become excessive so that additional instances can share the load. $U_{\text{low}}$ is the scale-in utilization threshold (default 20\%), used to reclaim instances when GPU utilization is insufficient. The scheduler runs in a control loop. Lines~1--2 initialize the instance counts $n$ and a history buffer $H$. Lines~3--5 periodically collect service metrics (utilization $u_s$, queue length $q_s$, average queueing delay $d_s$) and update $H$. Lines 6–10 implement the proactive adjustment mechanism. We first identifying the most frequent workload in $H$. Once a change is detected, the predictor $\hat{g}(\cdot)$ estimates the target instance counts based on features from the recent trace. Following the allocation update, the loop skips subsequent reactive logic to prevent interference, advancing immediately to the next iteration. Lines~11--17 implement the scheduling strategy. We trigger a scale-out operation only when the service GPU utilization reaches saturation, the queue length grows significantly, and the average queueing delay is increasing. Conversely, we initiate a scale-in operation when the utilization drops to a low level and the queue is empty, indicating that the pending workload has been cleared. Line~18 sleeps for $\Delta$ before the next iteration.

\section{Implementation}


\subsection{Network Communication}
\prjname{} employs ZeroMQ (ZMQ)~\cite{zeromq} for inter-stage signaling and the Mooncake Transfer Engine~\cite{mooncake_te_doc} for high-efficiency data transmission. By strictly separating the control plane (task metadata) from the data plane (intermediate results), the architecture effectively mitigates head-of-line blocking caused by large messages, thereby enhancing system stability under high-concurrency scenarios.

To ensure robust and efficient transmission, ZeroMQ sockets operate in an asynchronous non-blocking mode managed by a unified polling mechanism. The system incorporates resilience features such as exponential backoff retries to handle connection jitter and timeouts gracefully. Furthermore, \prjname{} exposes configurable parameters—including high-water marks and buffer sizes—enabling fine-grained optimization to balance throughput and latency across diverse hardware and workload patterns.

\subsection{Decentralized Queue Scheduling}\label{sec:decentral-queue}
\begin{figure}[t]
  \centering
  \includegraphics[width=0.9\linewidth]{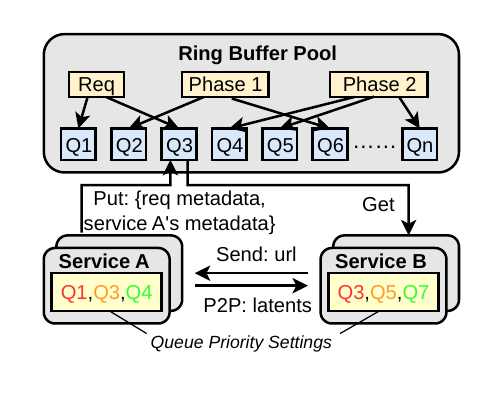}
  \caption{Decentralized queue scheduling.}
  \label{fig:decentralized-queue}
\end{figure}

\prjname{} employs a decentralized queue scheduling mechanism to meet the low-latency demands of generative inference. Successive stages are decoupled via RDMA-backed metadata queues using a producer--consumer pattern. This approach circumvents the bottlenecks and single-point-of-failure risks inherent to centralized queues, thereby improving multi-node throughput and robustness.

We split each request into two distinct objects: control plane (metadata) and data plane (payload). This module primarily focuses on providing low-latency request scheduling and instance coordination. As illustrated in Figure~\ref{fig:decentralized-queue}, the request scheduling path transmits only lightweight request metadata, writing these payloads into pre-registered RDMA buffer queue slots, while large-scale intermediate latents are accessed separately via Mooncake Transfer Engine. Leveraging fixed-length metadata, the system achieves $
O(1)$ queue operations, reducing memory copies and protocol overhead to improve stage-transition efficiency. We utilize RDMA Fetch-and-Add (FAA) atomic operations for lock-free queue concurrency control and leverage RDMA one-sided read/write verbs for producer--consumer queue access, replacing the traditional ZMQ push--pull model to provide a lower-latency communication channel.

To construct these RDMA-backed metadata queues, the Controller selects a small set of machines for each stage to host circular buffers and disseminates their addresses to instances. Each instance maintains a local \textit{queue table} that records the buffer locations for all stages. When accessing a given stage queue, it preferentially chooses the buffer with lower network latency to reduce communication overhead.

Each stage operates as a consumer of its upstream and a producer for its downstream, exchanging fixed-size metadata via a shared queue. This design decentralizes scheduling decisions, eliminating the global lock contention and control-plane congestion of centralized approaches. Additionally, on-demand consumption from shared queues allows for fine-grained load control without complex central coordination.

Furthermore, the decentralized queue scheduling significantly enhances scalability. To handle escalating loads, the system supports horizontal scaling of consumer instances at bottleneck stages, allowing dynamic queue subscription without reconfiguring centralized logic. Additionally, queue-level backpressure prevents congestion propagation: when downstream queues near capacity, upstream production is automatically rerouted to maintain overall pipeline stability.

\subsection{Zero-Copy and Batched Messaging}

To mitigate data movement overhead, \prjname{} adopts a zero-copy transmission strategy within the data plane. Memory is allocated directly on the GPU device memory, and inter-GPU tensor transfers are facilitated via the \textit{GDirect} mechanism~\cite{nvidia-gpudirect-rdma}. This approach substantially alleviates CPU overhead associated with memory copying, thereby improving resource utilization in high-throughput scenarios.

To optimize the transmission path, the system employs a message batching mechanism. By aggregating multiple small messages within a brief time window and transmitting them as a single batch, this approach effectively amortizes the overhead associated with system calls and per-message headers. The batching policy incorporates a dual-trigger approach based on both size and time thresholds: transmission is triggered immediately when the accumulated payload reaches the batch size limit or when the waiting time exceeds the timeout threshold. This design strikes a balance between throughput enhancement and tail-latency control.

\subsection{Fault Tolerance}

For fault detection, \prjname{} employs timeout-based strategies that enable communicating peers to promptly detect node disconnections, link anomalies, and processing stalls. Each request carries a unique request ID for end-to-end tracing and retry deduplication, preventing duplicate execution during failure recovery.

At the fault recovery level, the system adopts automatic reconnection with bounded retry for transient network failures, and triggers instance reclamation and reallocation for prolonged unavailability. Given the stateless nature of our services, failed instances can be directly substituted. Moreover, requests are decoupled from specific nodes and can be rerouted to any operational instance following a timeout, which significantly enhances system elasticity.

\section{Evaluation}\label{sec:evaluations}

In this section, we evaluate \prjname{} to answer the following questions:
\begin{itemize}
    \item Does \prjname{} preserve generation quality compared to baselines (\S\ref{sec:eval-quality})?
    \item How does \prjname{} improve end-to-end latency under different models, and deployment configurations (\S\ref{sec:eval-latency})?
    \item How scalable is \prjname{} as we scale out the distributed deployment (\S\ref{sec:eval-scalability})?
    \item How robust is \prjname{} to workload heterogeneity (\S\ref{sec:eval-sensitivity})?
    \item How does \prjname{} behave under different instance ratios and elastic scheduling policies (\S\ref{sec:eval-elastic})?
    \item How efficiently does \prjname{} utilize resources (\S\ref{sec:eval-utilization})?
\end{itemize}

\subsection{Experiment Setup}\label{sec:eval-setup}
\textbf{Testbed.} We run experiments on three two-server clusters: (1) two machines, each equipped with 8\,$\times$\,NVIDIA A10 GPUs; (2) two machines, each equipped with 8\,$\times$\,RTX~4090 GPUs; and (3) two machines, each equipped with 8\,$\times$\,NVIDIA H100 GPUs. All machines have identical CPU and memory configurations: two Intel Xeon Platinum 8369B CPUs (128 logical threads in total) and 502~GiB of DRAM. Each machine is connected via 100~Gbps RDMA.

\noindent\textbf{Models.} We evaluate \prjname{} on two representative generative AI workloads that emphasize visual synthesis. The first is \textit{Wan2.2}~\cite{wan22}, a state-of-the-art diffusion model. This workload exercises the complete DiT pipeline, including VAE encoding/decoding and large-scale latent tensor transfers, providing a rigorous test for our disaggregated architecture under high-bandwidth, synchronous dependencies. The second is \textit{Qwen-Image-2512}~\cite{qwen-image-2512}, a large-scale image generation model with 25.12 billion parameters. Although it generates images rather than videos, its massive model size (exceeding the memory capacity of a single A100 or consumer-grade GPU) and its adoption of a diffusion transformer backbone make it an ideal workload for stressing memory disaggregation and stage-level pipeline parallelism. Unless otherwise stated, we use the official releases and their default inference configurations, varying request parameters (e.g., diffusion steps or generation length) as specified in each experiment.

\noindent\textbf{Baselines.} We compare against the monolithic deployment of LightX2V as our baseline. To ensure a fair comparison, we strictly match the total number of GPUs between \prjname{} and the baseline across all experiments. In multi-GPU inference scenarios, we employ tensor parallelism to optimize the baseline's multi-GPU performance, thereby establishing a strong baseline for evaluating the benefits of \prjname{}.

\subsection{Generation Quality}\label{sec:eval-quality}

\begin{figure}[t]
    \centering
    \includegraphics[width=0.232\linewidth]{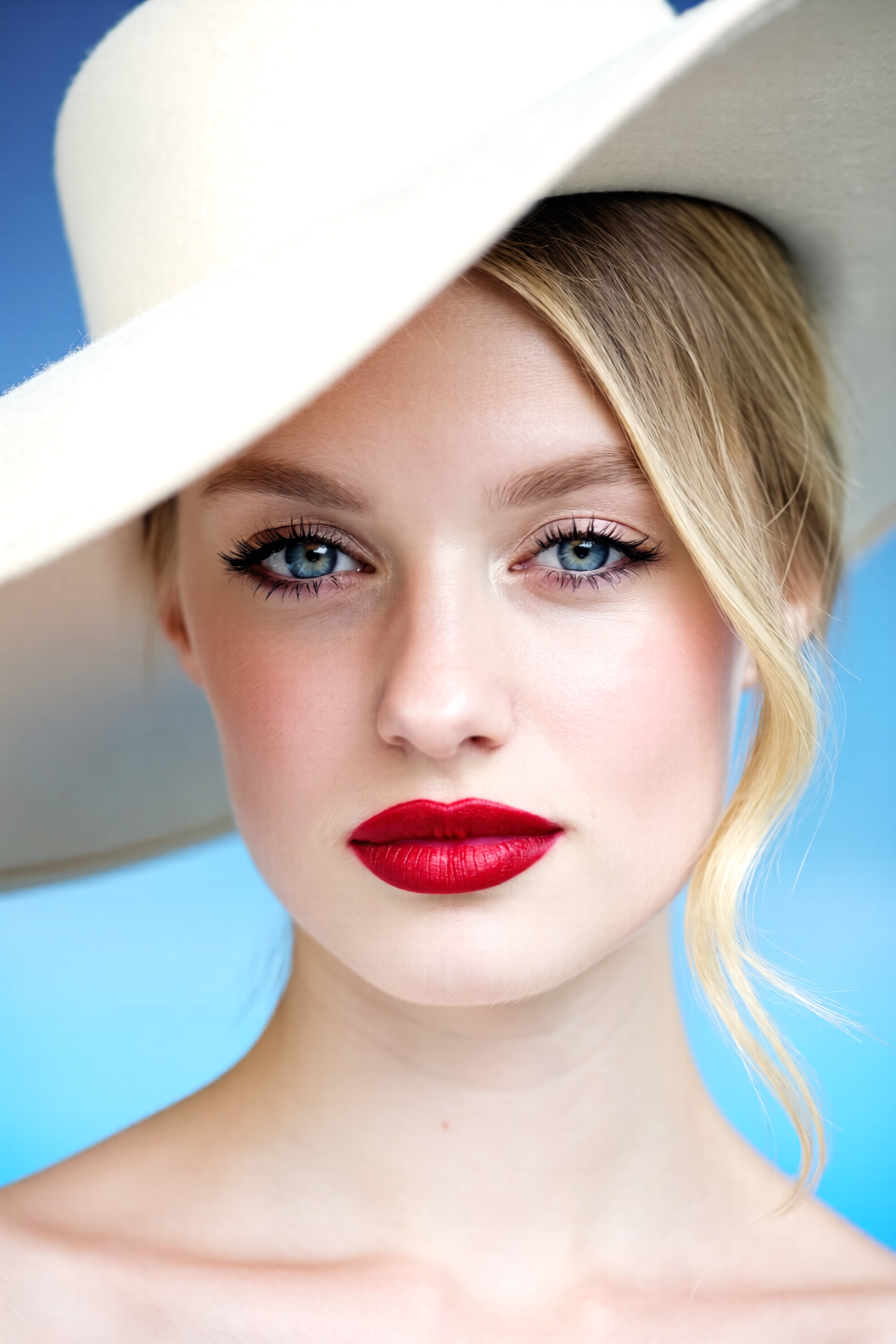}\hfill
    \includegraphics[width=0.261\linewidth]{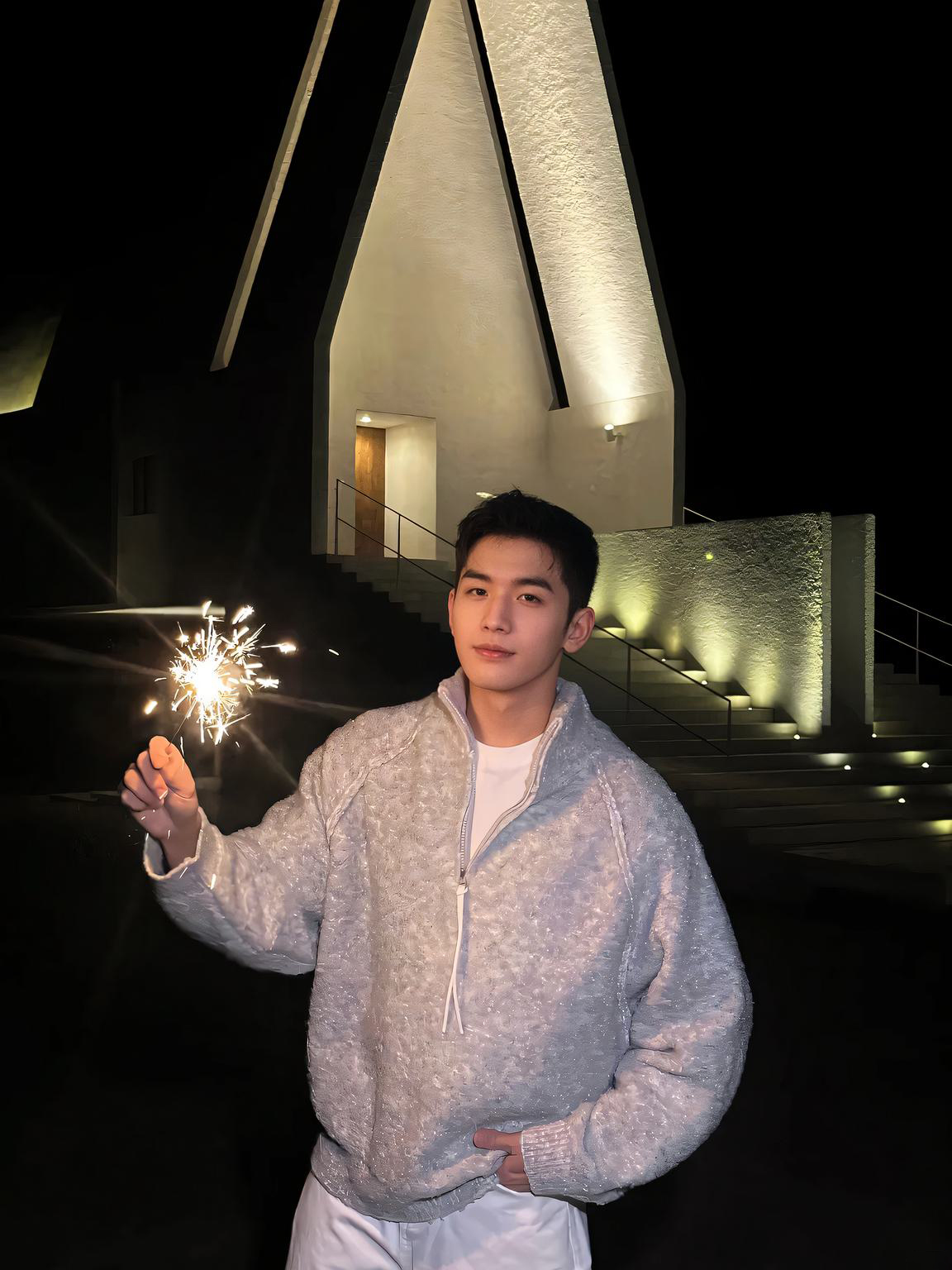}\hfill
    \includegraphics[width=0.262\linewidth]{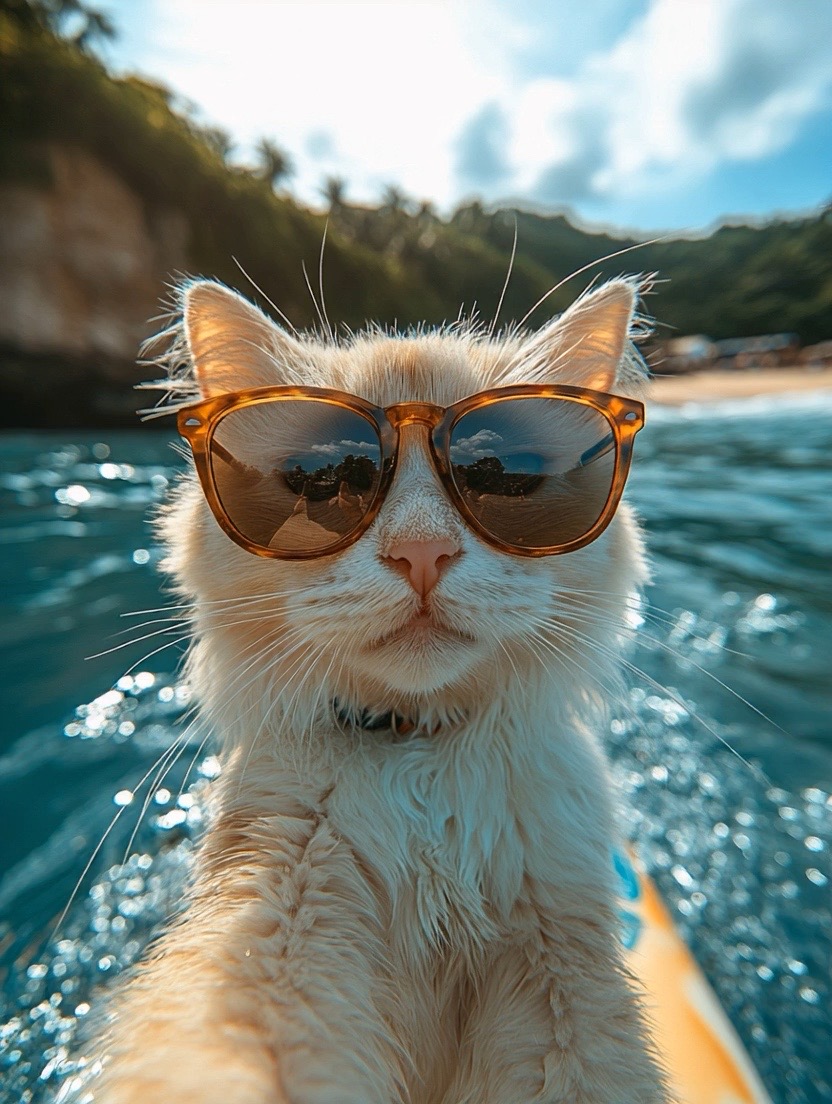}\hfill
    \includegraphics[width=0.233\linewidth]{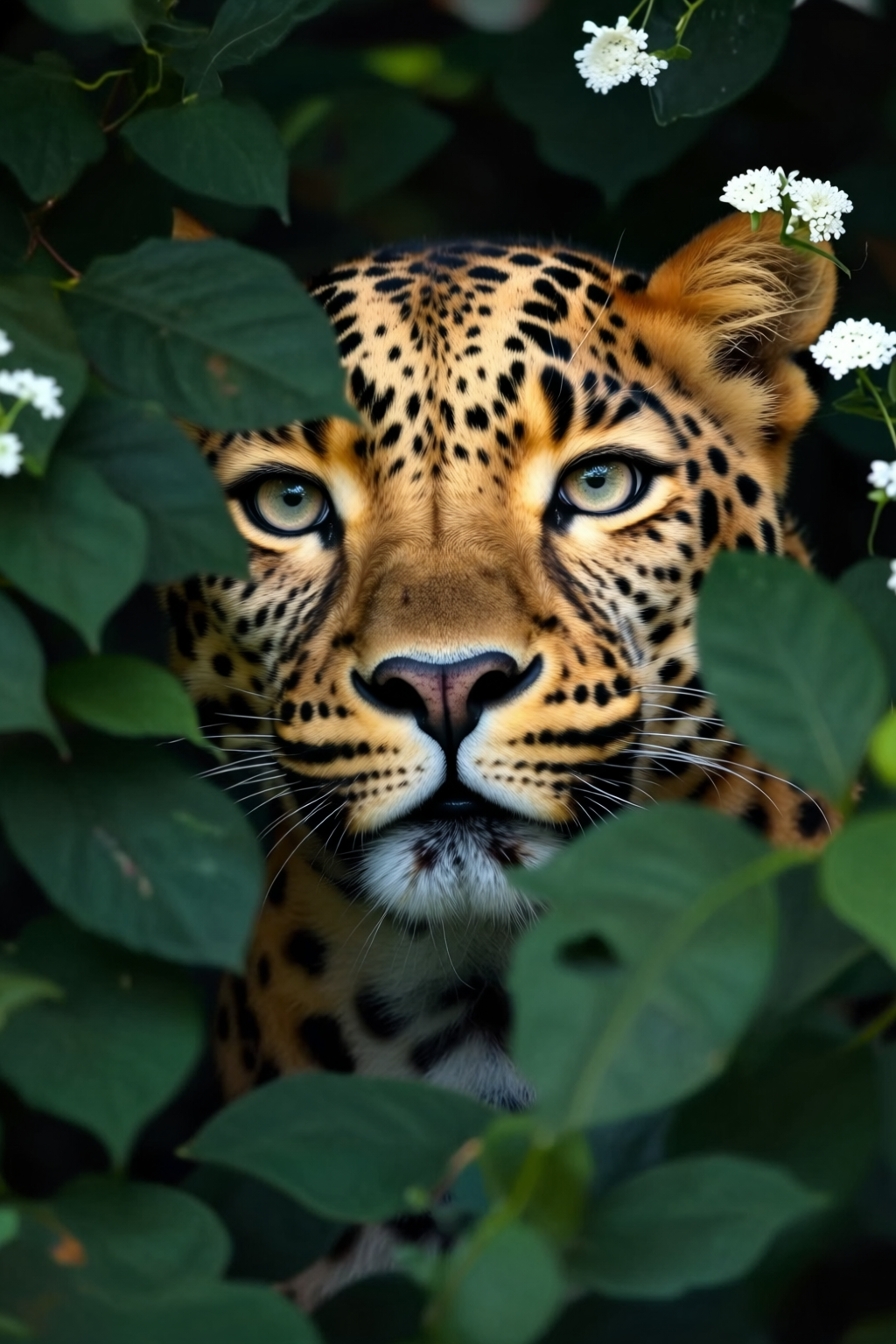}
    \caption{Examples of images generated by \prjname{}.}
    \label{fig:fig-gen}
\end{figure}

\begin{table}[t]
    \centering
    \caption{Generation-quality metrics. In each cell, the score is the baseline result, and \prjname{} achieves scores consistent with the baseline. (SC: subject consistency; BC: background consistency; AQ: aesthetic quality; IQ: imaging quality.) }
    \label{tab:quality-metrics}
    \small
    \begin{tabular}{lcccc}
        \toprule
        \textbf{Workload} & \textbf{SC} & \textbf{BC} & \textbf{AQ} & \textbf{IQ} \\
        \midrule
        T2V 50-step & 0.960 & 0.938 & 0.655 & 0.595 \\
        I2V 40-step & 0.993 & 0.971 & 0.673 & 0.684 \\
        I2V 8-step & 0.993 & 0.969 & 0.667 & 0.678 \\
        I2V 4-step & 0.993 & 0.967 & 0.664 & 0.676 \\
        I2V 1-step & 0.967 & 0.955 & 0.632 & 0.721 \\
        \bottomrule
    \end{tabular}
\end{table}

We validate that \prjname{}’s stage disaggregation does not compromise generation quality. Since \prjname{} splits the end-to-end pipeline into Encoder/Transformer/Decoder stages and transfers intermediate tensors across stage boundaries, we want to confirm that the produced outputs remain indistinguishable from the monolithic baseline.
We run the same prompts and random seeds on a single node with 8 GPUs and 16 GPUs, respectively, and compare against the monolithic baseline under the same GPU budgets. We use a resolution of 832$\times$480 and generate 81 frames for each request.

Figure~\ref{fig:fig-gen} shows real examples generated by \prjname{}, and the results are visually identical to those produced by LightX2V.
Table~\ref{tab:quality-metrics} reports the quality scores for T2V and I2V. The results show that \prjname{} produces quality scores consistent with the monolithic baseline, indicating that stage separation and inter-stage transfers do not affect the final outputs.
In addition to end-to-end quality metrics, we perform a tensor-level correctness validation by attaching hash checks to transmitted tensors. This verifies that tensors received by downstream stages exactly match those produced by upstream stages, ruling out data corruption or unintended numerical deviations during transfer.

\begin{figure}[t]
    \centering
    \captionsetup[subfloat]{font=small}
    \subfloat[Wan2.2 (A10), I2V 4-step.\label{fig:latency-cdf-wan}]{\includegraphics[width=\linewidth]{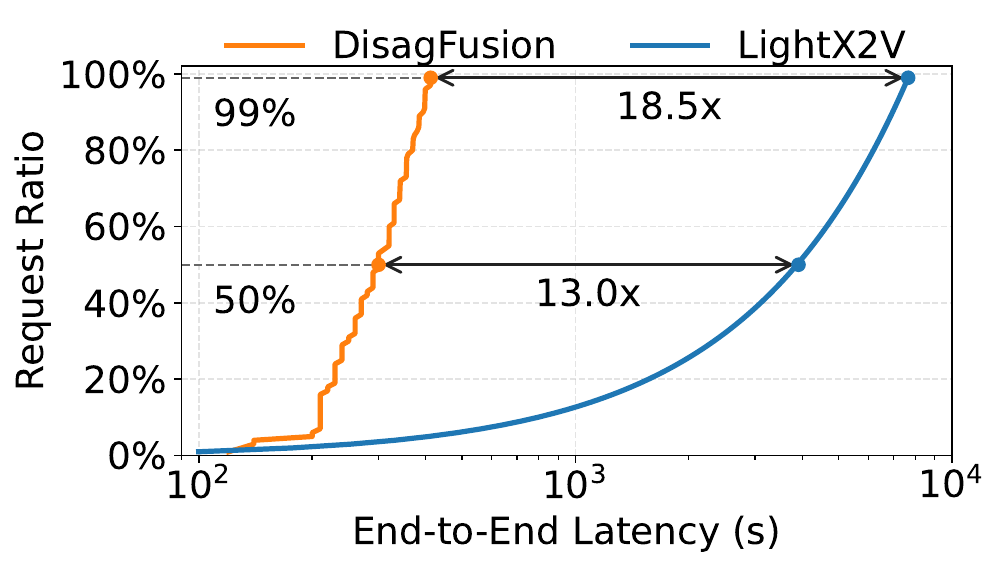}}\hfill
    \subfloat[Qwen2512 (4090), T2I 8-step.\label{fig:latency-cdf-qwen}]{\includegraphics[width=\linewidth]{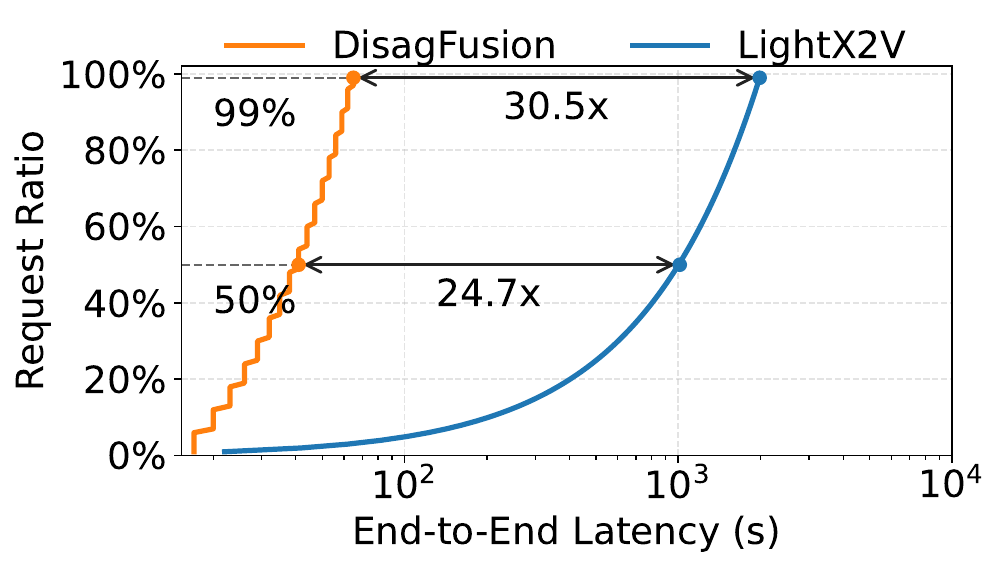}}
    \caption{End-to-end latency comparison between LightX2V and \prjname{} in serving requests.}
    \label{fig:latency-cdf}
\end{figure}

\subsection{End-to-End Latency}\label{sec:eval-latency}

\begin{figure*}[t]
    \centering
    \captionsetup[subfloat]{font=small}
    \includegraphics[width=0.6\textwidth]{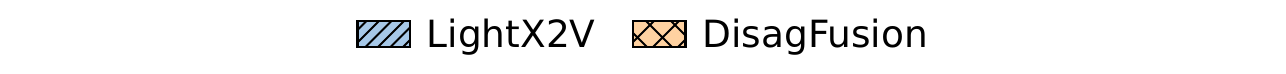}\\
    \vspace{-1em}
    \subfloat[Wan2.2 (A10), T2V 50-step.\label{fig:scalability-wan22-a10-50step-eval}]{%
        \includegraphics[width=0.32\textwidth]{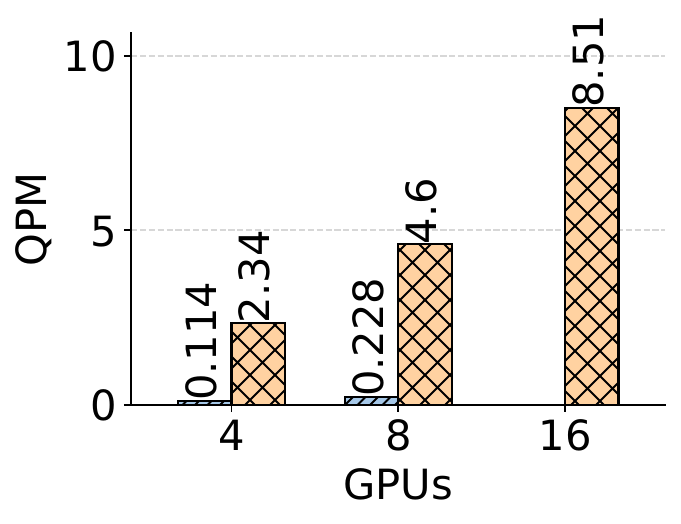}%
    }\hfill
    \subfloat[Wan2.2 (A10), I2V 4-step.\label{fig:scalability-wan22-a10-4step-eval}]{%
        \includegraphics[width=0.32\textwidth]{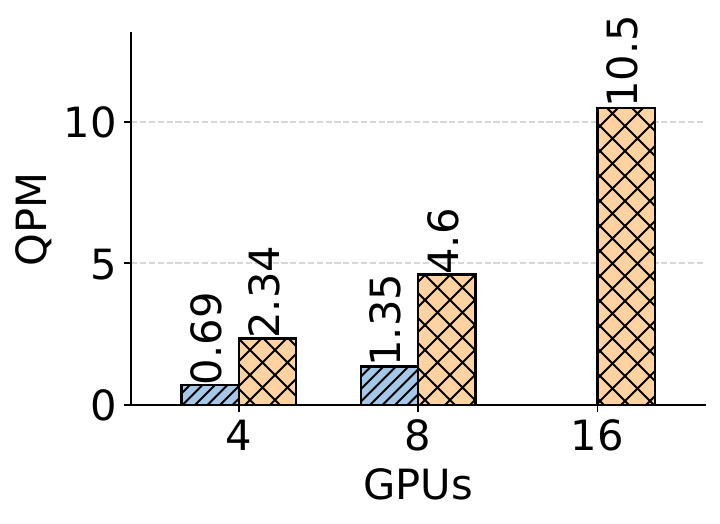}%
    }\hfill
    \subfloat[Qwen2512 (4090), T2I 8-step.\label{fig:scalability-qwen-4090-8step-eval}]{%
        \includegraphics[width=0.32\textwidth]{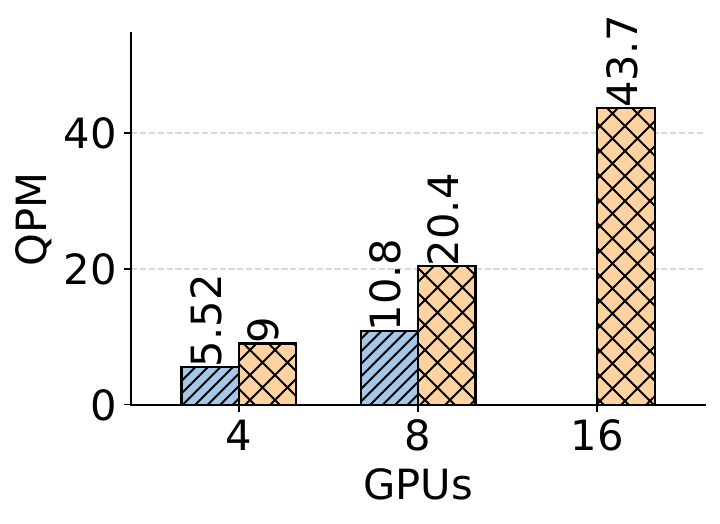}%
    }
    \caption{Comparison of scalability for LightX2V and \prjname{} under different workloads.}
    \label{fig:scalability-eval}
\end{figure*}

In this experiment, we evaluate the end-to-end latency of \prjname{} and LightX2V using \textit{Wan2.2}~\cite{wan22} and \textit{Qwen2512}~\cite{qwen-image-2512} models. We run the Wan2.2 experiments on eight A10 GPUs and the Qwen2512 experiments on eight RTX 4090 GPUs. As illustrated by the CDF curves in Figure~\ref{fig:latency-cdf}, \prjname{} consistently exhibits a significant leftward shift compared to LightX2V across both models, indicating superior latency performance.

Specifically, for the \textit{Wan2.2} model (Figure~\ref{fig:latency-cdf-wan}), the 50\% and 99\% latencies of \prjname{} are 13.0$\times$ and 18.5$\times$ lower than those of LightX2V, respectively. The performance gains are even more pronounced with the large-scale \textit{Qwen2512} model (Figure~\ref{fig:latency-cdf-qwen}), where \prjname{} achieves median and tail latencies that are 24.7$\times$ and 30.5$\times$ lower than LightX2V.
This substantial improvement stems from two key factors: (i) \prjname{} shortens per-request inference time by disaggregating computation into stages, avoiding the blocking GPU load/unload operations required by the monolithic baseline; and (ii) \prjname{} employs pipelined execution to overlap requests across stages, thereby minimizing queue buildup. In contrast, LightX2V processes requests synchronously and serially, leading to lower GPU utilization and higher queueing delays, particularly under the heavy load of large models.

\subsection{Distributed Scalability}\label{sec:eval-scalability}

To evaluate \prjname{}'s scalability, we measure the throughput of \prjname{} and the LightX2V monolithic baseline with 4, 8, and 16 GPUs. For each test, we send 30 identical requests. The results are shown in Figure~\ref{fig:scalability-eval}.

Both systems exhibit near-linear scaling within the supported range; however, \prjname{} consistently delivers superior performance. A critical limitation of the LightX2V baseline is its single-node deployment architecture, which restricts it to a maximum of 8 GPUs and prevents it from scaling to the 16-GPU multi-node configuration. In contrast, \prjname{} successfully scales across all settings.
Specifically, in the T2V 50-step workload (Figure~\ref{fig:scalability-wan22-a10-50step-eval}), \prjname{} achieves throughputs of 2.34, 4.6, and 8.51 QPM on 4, 8, and 16 GPUs, respectively. This significantly outperforms LightX2V, delivering approximately 20.5$\times$ and 20.3$\times$ higher throughput on 4 and 8 GPUs. Similarly, for the I2V 4-step task (Figure~\ref{fig:scalability-wan22-a10-4step-eval}), \prjname{} scales from 2.34 to 10.5 QPM, surpassing the baseline by factors of 3.4$\times$ and 7.7$\times$.
In the evaluation using the Qwen2512 (4090) model (Figure~\ref{fig:scalability-qwen-4090-8step-eval}), \prjname{} successfully scales to 16 GPUs, achieving 43.7 QPM---more than double its 8-GPU performance and 4.0$\times$ faster than the baseline on 8 GPUs. This robust scalability stems from \prjname{}'s architecture, where independent instances fully utilize GPU resources without cross-instance dependencies, allowing aggregate throughput to grow directly with the number of GPUs.

\begin{figure}[t]
    \centering
    \includegraphics[width=\linewidth]{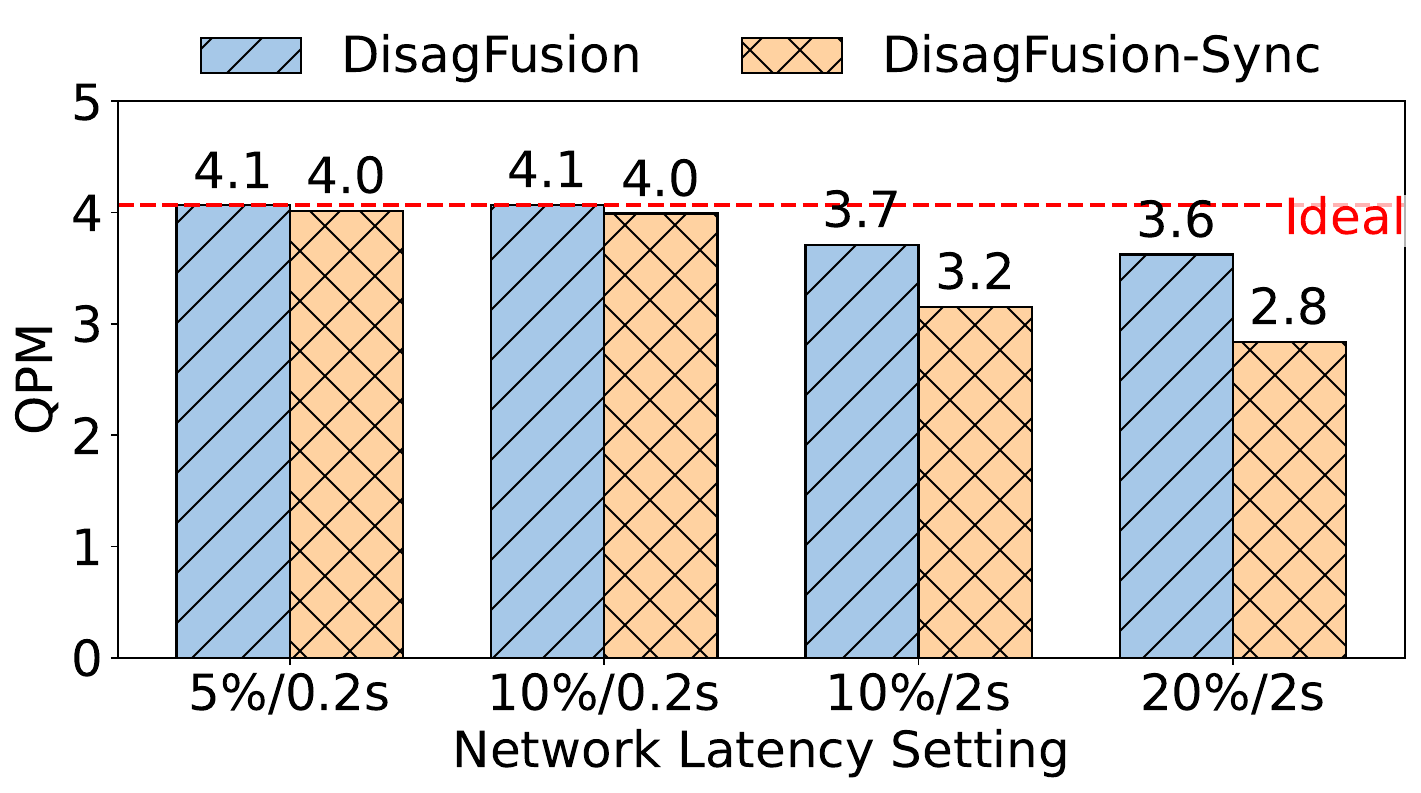}
    \caption{Network latency comparison between \prjname{} and \prjname{}'s synchronous variant.}
    \label{fig:network-latency-eval}
\end{figure}

\subsection{Robustness Analysis}\label{sec:eval-sensitivity}
Disaggregated serving is inherently sensitive to network conditions because inter-stage tensor transfers lie on the critical path. To evaluate \prjname{}'s robustness against unpredictable network behavior, we simulate four jitter patterns commonly observed in production clusters (Figure~\ref{fig:network-latency-eval}): (1) stable network, where each transfer has a 5\% probability of incurring an additional 0.2\,s delay; (2) mild jitter, 10\%/0.2\,s; (3) moderate jitter, 10\%/2\,s; and (4) severe jitter, 20\%/2\,s.

Under mild jitter, both synchronous and asynchronous designs remain stable. However, as jitter becomes more severe, the synchronous baseline suffers drastic throughput drops---22.5\% under moderate jitter and 30.3\% under severe jitter---because the upstream stage blocks until the downstream acknowledges receipt, turning every network delay into GPU idle time. In contrast, \prjname{} limits the degradation to 8.8\% and 11.0\%, respectively. By decoupling stages via asynchronous queues, \prjname{} overlaps the communication of one request with the computation of another; transient jitter is therefore absorbed by queue buffering rather than propagated as pipeline bubbles, making the system substantially less sensitive to network fluctuations.

\begin{figure}[t]
    \centering
    \captionsetup[subfloat]{font=small}
    \subfloat[Real-time throughput under varying request parameters.\label{fig:realtime-thrput-p-eval}]{%
        \includegraphics[width=\linewidth]{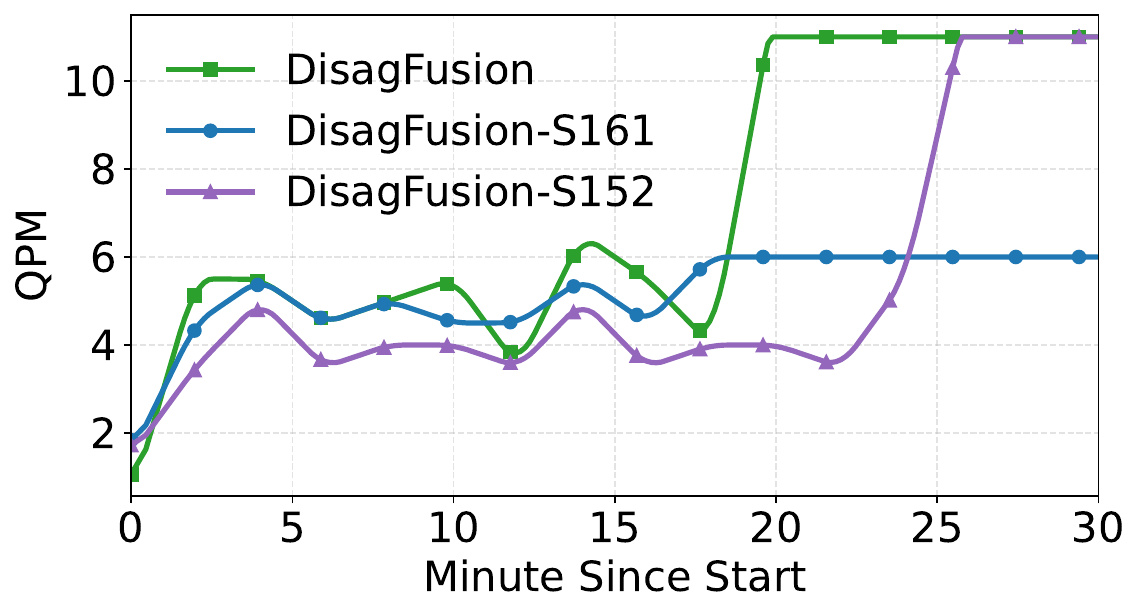}%
    }\\
    \subfloat[Real-time throughput under varying request rates.\label{fig:realtime-thrput-r}]{%
        \includegraphics[width=\linewidth]{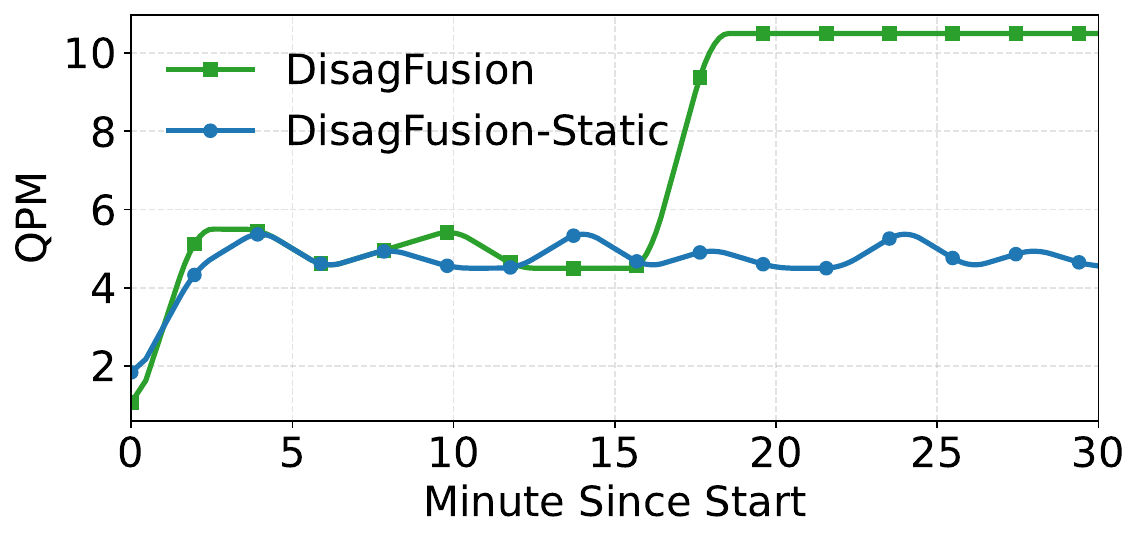}%
    }
    \caption{Real-time throughput performance under dynamic workloads.}
    \label{fig:realtime-thrput}
\end{figure}

\subsection{Instance Ratio \& Elastic Scheduling}\label{sec:eval-elastic}
To study different strategies under dynamic load, we measure the real-time throughput of \prjname{} and the monolithic baseline under two workload traces that vary request parameters and request rates. For both traces, the first 15 minutes use 4-step requests with a fixed arrival rate of 0.1\,req/s. After 15 minutes, the \textit{parameter-varying} trace switches the incoming requests to 1-step, while the \textit{rate-varying} trace increases the arrival rate to 0.2\,req/s. Figure~\ref{fig:realtime-thrput-p-eval} and Figure~\ref{fig:realtime-thrput-r} report the results, respectively.

In the parameter-varying trace, we compare \prjname{}'s dynamic scheduling against fixed instance allocations (1:6:1 and 1:5:2).
During the first 15 minutes (4-step requests), the DiT stage is the bottleneck for both fixed settings. The 1:6:1 allocation (\prjname{}-S161) achieves a throughput of 4.9 QPM, outperforming the 1:5:2 allocation (\prjname{}-S152), which is more severely constrained at 4.0 QPM. However, after the switch to 1-step requests (t > 15 min), the bottleneck shifts. For the 1:6:1 setting, the bottleneck moves to the Decoder, capping throughput at 6.2 QPM. In contrast, the 1:5:2 setting becomes Encoder-bottlenecked, allowing it to reach 11.0 QPM.
\prjname{} automatically adapts to these changes: it implicitly aligns with the optimal 1:6:1 configuration in the first window and switches to the 1:5:2 configuration in the second, thereby sustaining the maximum possible throughput throughout the trace.

In the request-rate-varying trace, we evaluate \prjname{}'s dynamic scale-out capability. During the first 15 minutes, the request rate is low (0.1req/s), allowing the initial 1:6:1 allocation to serve requests without queueing. However, when the rate doubles to 0.2req/s after 15 minutes, the system with only 8 GPUs becomes saturated, causing the request queue to build up.
To address this, \prjname{}'s elastic scheduling detects the backpressure and automatically provisions an additional 8-GPU machine, shifting the instance ratio to 1:13:2. Under this expanded configuration, the throughput bottleneck shifts to the DiT stage, raising the system's capacity to 10.5 QPM. Consequently, \prjname{} successfully absorbs the increased load and stabilizes at the new, higher throughput, whereas a static configuration suffers from unbounded queuing delays.

\begin{figure}[t]
    \centering
    \includegraphics[width=\linewidth]{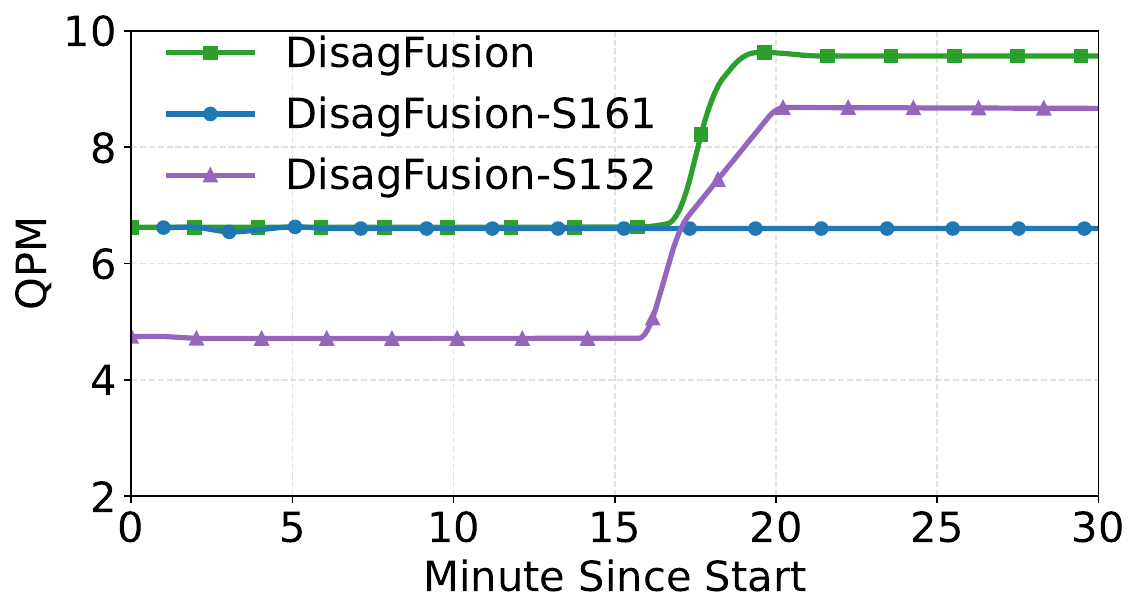}
    \caption{Real-time throughput under varying request parameters on the H100 cluster.}
    \label{fig:realtime-thrput-p-h100}
\end{figure}

Furthermore, we evaluate the system on the H100 cluster under varying request parameter workloads. As shown in Figure~\ref{fig:realtime-thrput-p-h100}, \prjname{} consistently achieves the highest throughput, maintaining 6.72 QPM in the first phase and scaling up to 9.62 QPM in the second. In contrast, static baselines fail to adapt efficiently. The 1:6:1 allocation (\prjname{}-S161) plateaus at 6.64 QPM, unable to exploit the increased request rate. Meanwhile, the 1:5:2 allocation (\prjname{}-S152) suffers from a severe bottleneck in the first phase (4.75 QPM) and, despite recovering to 8.67 QPM later, still lags behind \prjname{}. These results confirm that \prjname{}'s dynamic scheduling is essential for maximizing hardware utilization across varying workloads.

\subsection{Resource Utilization}\label{sec:eval-utilization}

\begin{figure}[t]
    \centering
    \includegraphics[width=\linewidth]{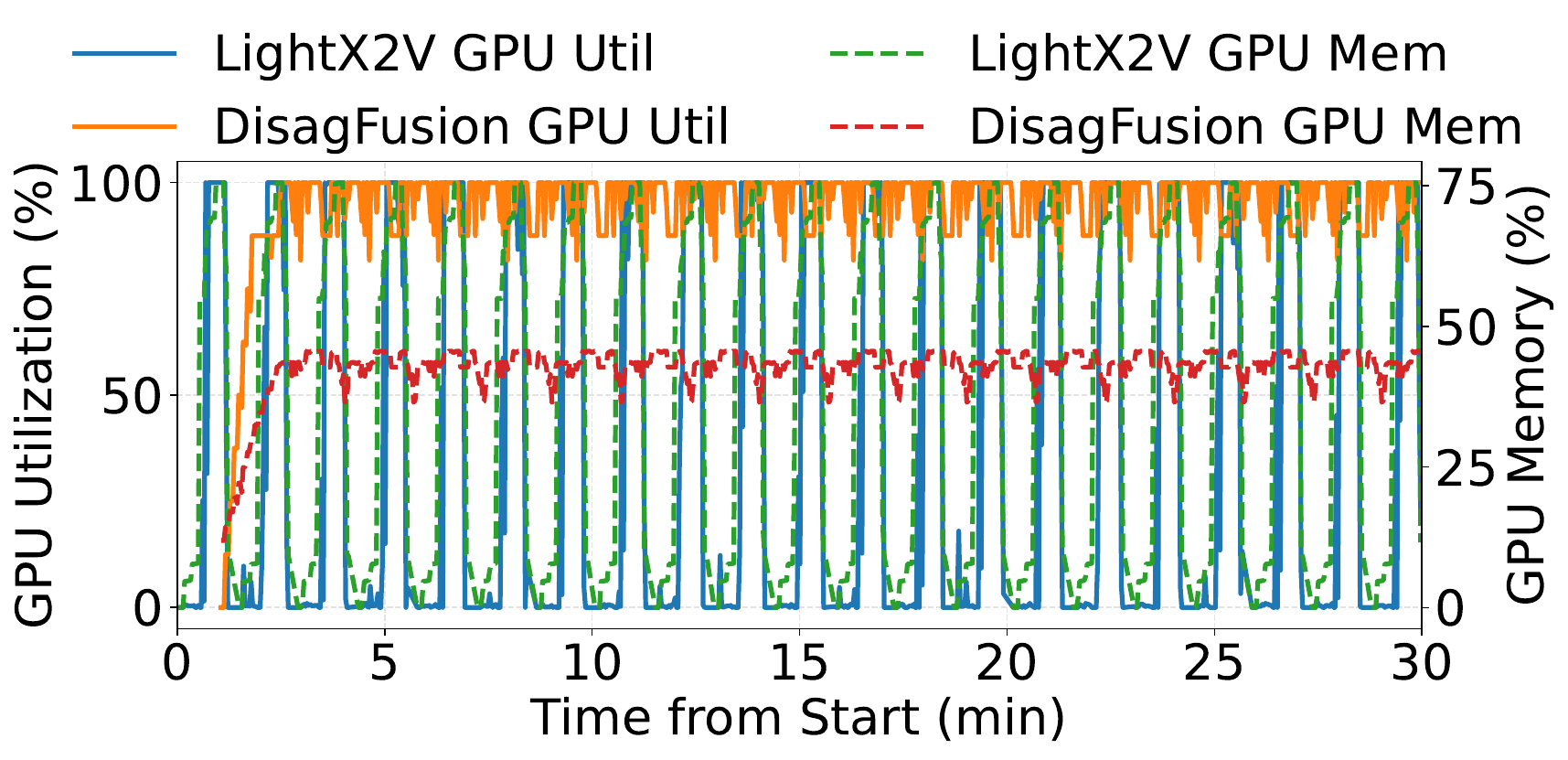}
    \caption{Comparison of GPU utilization and memory footprint under different deployment strategies (LightX2V and \prjname{}).}
    \label{fig:gpu-util-compare}
\end{figure}

We next evaluate how efficiently \prjname{} utilizes GPU resources compared to the monolithic LightX2V baseline. We run a 30-minute continuous serving experiment using 4-step Wan2.2 requests, and record per-GPU utilization and memory footprint throughout the run under the same total GPU budget.

\noindent\textbf{Sustained and smooth utilization.} Figure~\ref{fig:gpu-util-compare} compares the GPU utilization time series. The monolithic baseline exhibits pronounced utilization oscillations: GPUs are often idle while the service is waiting for CPU-side orchestration, intermediate transfers, or model (un)loading, and the bursty execution leads to frequent under-utilized periods. In contrast, \prjname{} maintains consistently higher and smoother utilization across GPUs. This is because stage disaggregation enables (i) better intra-stage load balance (each stage runs a homogeneous kernel mix) and (ii) pipelined overlap across requests, so different stages can remain busy even when individual requests experience transient stalls.

\noindent\textbf{Reduced and stable memory pressure.} We also track GPU memory usage during the 30-minute run. The baseline must keep more components resident simultaneously to execute the end-to-end model within a single process, leading to a higher and more volatile memory footprint and leaving less headroom for batching and concurrency. In contrast, \prjname{} separates model components across distinct stages and machines, ensuring that each GPU maintains only the parameters and activations for its assigned stage. This alleviates peak per-GPU memory pressure and stabilizes memory footprints over time, thereby enhancing robustness under concurrent workloads.

\section{Related Works}

\textbf{Diffusion serving and diffusion systems.} Recent research efforts have emerged to systematically organize and enhance the serving of diffusion models.~\cite{lin2025understanding} provides a production-driven analysis of diffusion serving, highlighting the impact of workload dynamics and scheduling decisions on efficiency. On the model side, diffusion-based generation has rapidly evolved from early diffusion and text-to-video systems~\cite{ho2022videodiffusion,singer2022makeavideo,villegas2022phenaki} to recent latent diffusion and adaptation techniques~\cite{blattmann2023videoldm,stability2023stablevideodiffusion,guo2023animatediff,wu2023tuneavideo,ho2022imagenvideo}. On the systems side, research initiatives focus on enhancing the throughput and latency of diffusion transformers through optimizations at the kernel, memory, and execution levels (e.g., DiT-Serve~\cite{luodit}). Alternatively, other approaches aim to reduce end-to-end overhead via the integration of lightweight serving components (e.g., SwiftDiffusion~\cite{li2024swiftdiffusion} and DiffServe~\cite{wang2024diffserve}). Recent systems also study workflow-aware serving when a diffusion pipeline is augmented with many adapters (e.g., LegoDiffusion~\cite{legodiffusion2026} and Katz~\cite{li2025katz}) and optimize mask-based image editing pipelines via caching and scheduling (e.g., FlashPS~\cite{jiang2026flashps}). Production serving stacks are evolving to integrate diffusion pipelines as native components, thereby unifying multimodal serving capabilities (e.g., SGLang Diffusion~\cite{sglang2025diffusion_blog,sglang_diffusion_docs}). Meanwhile, specialized frameworks such as LightX2V~\cite{lightx2v} offer tailored optimizations to streamline the deployment of generation models. Orthogonally, distributed inference frameworks such as DistriFusion~\cite{li2024distrifusion} leverage multi-GPU parallelism to distribute diffusion inference, thereby facilitating high-resolution image generation. In contrast to these approaches, \prjname{} focuses on a disaggregated, stage-separated architecture tailored specifically for diffusion. This work characterizes the unique performance bottlenecks arising from inter-stage tensor transfers within this architectural paradigm.

\textbf{Disaggregated serving for LLMs/multimodal models.} Disaggregated serving has also been extensively studied for LLMs and multimodal models, where the goal is to decouple heterogeneous phases and improve goodput under dynamic load. Prior systems disaggregate computation and memory (e.g., Mooncake~\cite{qin2024mooncake}) or split prefill and decoding to enable better multiplexing and resource provisioning (e.g., Orca~\cite{yu2022orca}, AlpaServe~\cite{li2023alpaserve}, and DistServe~\cite{zhong2024distserve}). On one hand, frameworks target the computational layer via continuous batching and advanced scheduling (e.g., PagedAttention~\cite{vllm2023pagedattention} and Sarathi-Serve~\cite{agrawal2024sarathiserve}). On the other hand, memory-centric approaches address memory constraints by exploiting model offloading and heterogeneous memory solutions to support large-scale generative models~\cite{sheng2023flexgen}. These designs demonstrate the general benefits of disaggregation, but they target token-level generation with KV-cache management as the dominant concern. \prjname{} addresses a different setting: diffusion generation has a natural Encoder--Transformer--Decoder structure with highly imbalanced stage costs, and stage separation makes network jitter and inter-stage backpressure first-order performance factors.

\section{Conclusion}

This paper presents \prjname{}, enabling asynchronous pipeline parallelism and elastic scheduling for disaggregated diffusion serving. By decoupling the pipeline into heterogeneous stages and introducing asynchronous pipeline parallelism, \prjname{} effectively addresses memory constraints and stage imbalance. Combined with a hybrid scheduling strategy, it achieves 3.4$\times$--20.5$\times$ throughput improvement and 18.5$\times$ reduction in latency compared to monolithic deployments, significantly enhancing serving efficiency.


\bibliographystyle{ACM-Reference-Format}
\normalem
\bibliography{sample-base}

@misc{lightx2v,
  title        = {LightX2V},
  howpublished = {GitHub repository},
  author       = {{ModelTC}},
  year         = {2025},
  url          = {https://github.com/ModelTC/LightX2V}
}

@article{qin2024mooncake,
  title={Mooncake: A kvcache-centric disaggregated architecture for llm serving},
  author={Qin, Ruoyu and Li, Zheming and He, Weiran and Cui, Jialei and Tang, Heyi and Ren, Feng and Ma, Teng and Cai, Shangming and Zhang, Yineng and Zhang, Mingxing and others},
  journal={ACM Transactions on Storage},
  year={2024},
  publisher={ACM New York, NY}
}

@misc{mooncake_te_doc,
  title        = {Mooncake Transfer Engine Design},
  howpublished = {Online documentation},
  author       = {{Mooncake Project}},
  year         = {2025},
  url          = {https://kvcache-ai.github.io/mooncake/docs/transfer-engine.html}
}

@misc{zeromq,
  title        = {ZeroMQ},
  howpublished = {Project website},
  author       = {{ZeroMQ Community}},
  year         = {2026},
  url          = {https://zeromq.org/}
}

@inproceedings{coppock2025lithos,
  title={LithOS: An operating system for efficient machine learning on GPUs},
  author={Coppock, Patrick H and Zhang, Brian and Solomon, Eliot H and Kypriotis, Vasilis and Yang, Leon and Sharma, Bikash and Schatzberg, Dan and Mowry, Todd C and Skarlatos, Dimitrios},
  booktitle={Proceedings of the ACM SIGOPS 31st Symposium on Operating Systems Principles},
  pages={1--17},
  year={2025}
}

@inproceedings{lin2025understanding,
  title={Understanding Diffusion Model Serving in Production: A Top-Down Analysis of Workload, Scheduling, and Resource Efficiency},
  author={Lin, Yanying and Wu, Shuaipeng and Luo, Shutian and Xu, Hong and Shen, Haiying and Ma, Chong and Shen, Min and Chen, Le and Xu, Chengzhong and Qu, Lin and others},
  booktitle={Proceedings of the 2025 ACM Symposium on Cloud Computing},
  pages={1--15},
  year={2025}
}

@misc{luodit,
  title        = {DiT-Serve: An Efficient Serving Engine for Diffusion Transformers},
  author       = {Luo, Michael and Hao, Aaron and Yan, Zhengxu and Cao, Chengkun and Nguyen, Quang Luong Nhat},
  year         = {2025},
  howpublished = {arXiv preprint},
  note         = {To appear},
}

@article{qiu2025modserve,
  title={Modserve: Scalable and resource-efficient large multimodal model serving},
  author={Qiu, Haoran and Biswas, Anish and Zhao, Zihan and Mohan, Jayashree and Khare, Alind and Choukse, Esha and Goiri, {\'I}{\~n}igo and Zhang, Zeyu and Shen, Haiying and Bansal, Chetan and others},
  journal={arXiv preprint arXiv:2502.00937},
  year={2025}
}

@misc{soprompts_sora_runway_pika,
  title        = {Sora vs Runway vs Pika: Comparison},
  howpublished = {Blog post},
  author       = {{SoPrompts}},
  year         = {2026},
  url          = {https://soprompts.com/blog/sora-vs-runway-vs-pika}
}

@misc{vidwave_pika_vs_sdv,
  title        = {Pika Labs vs Stable Diffusion Video: Quality Test Results},
  howpublished = {Blog post},
  author       = {{Vidwave}},
  year         = {2026},

  url          = {https://vidwave.ai/pika-labs-vs-stable-diffusion-video-quality-test-results}
}

@misc{vllm2023pagedattention,
  title        = {Efficient Memory Management for Large Language Model Serving with PagedAttention},
  author       = {Kwon, Woosuk and Li, Zhuohan and Zhuang, Siyuan and Sheng, Ying and Zheng, Lianmin and Yu, Cody Hao and Gonzalez, Joseph E. and Zhang, Hao and Stoica, Ion},
  year         = {2023},
  howpublished = {arXiv preprint arXiv:2309.06180},
  doi          = {10.48550/arXiv.2309.06180}
}

@inproceedings{yu2022orca,
  title     = {Orca: A Distributed Serving System for {Transformer-Based} Generative Models},
  author    = {Yu, Gyeong-In and Jeong, Joo Seong and Kim, Geon-Woo and Kim, Soojeong and Chun, Byung-Gon},
  booktitle = {16th USENIX Symposium on Operating Systems Design and Implementation (OSDI 22)},
  year      = {2022},
  pages     = {521--538},
  publisher = {USENIX Association}
}

@inproceedings{zhong2024distserve,
  title     = {{DistServe}: Disaggregating Prefill and Decoding for Goodput-optimized Large Language Model Serving},
  author    = {Zhong, Yinmin and Liu, Shengyu and Chen, Junda and Hu, Jianbo and Zhu, Yibo and Liu, Xuanzhe and Jin, Xin and Zhang, Hao},
  booktitle = {18th USENIX Symposium on Operating Systems Design and Implementation (OSDI 24)},
  year      = {2024},
  pages     = {193--210},
  publisher = {USENIX Association}
}

@inproceedings{li2023alpaserve,
  title     = {{AlpaServe}: Statistical Multiplexing with Model Parallelism for Deep Learning Serving},
  author    = {Li, Zhuohan and Zheng, Lianmin and Zhong, Yinmin and Liu, Vincent and Sheng, Ying and Jin, Xin and Huang, Yanping and Chen, Zhifeng and Zhang, Hao and Gonzalez, Joseph E. and Stoica, Ion},
  booktitle = {17th USENIX Symposium on Operating Systems Design and Implementation (OSDI 23)},
  year      = {2023},
  pages     = {663--679},
  publisher = {USENIX Association}
}

@misc{agrawal2024sarathiserve,
  title        = {Taming Throughput-Latency Tradeoff in LLM Inference with Sarathi-Serve},
  author       = {Agrawal, Amey and Kedia, Nitin and Panwar, Ashish and Mohan, Jayashree and Kwatra, Nipun and Gulavani, Bhargav S. and Tumanov, Alexey and Ramjee, Ramachandran},
  year         = {2024},
  howpublished = {arXiv preprint arXiv:2403.02310},
  doi          = {10.48550/arXiv.2403.02310}
}

@article{sheng2023flexgen,
  title   = {FlexGen: High-Throughput Generative Inference of Large Language Models with a Single GPU},
  author  = {Sheng, Ying and Zheng, Lianmin and Yuan, Binhang and Li, Zhuohan and Ryabinin, Max and Fu, Daniel Y. and Xie, Zhiqiang and Chen, Beidi and Barrett, Clark and Gonzalez, Joseph E. and Liang, Percy and {R{\'e}}, Christopher and Stoica, Ion and Zhang, Ce},
  journal = {arXiv preprint arXiv:2303.06865},
  year    = {2023},
  doi     = {10.48550/arXiv.2303.06865}
}

@article{ho2020ddpm,
  title   = {Denoising Diffusion Probabilistic Models},
  author  = {Ho, Jonathan and Jain, Ajay and Abbeel, Pieter},
  journal = {arXiv preprint arXiv:2006.11239},
  year    = {2020},
  doi     = {10.48550/arXiv.2006.11239}
}

@inproceedings{rombach2022ldm,
  title     = {High-Resolution Image Synthesis with Latent Diffusion Models},
  author    = {Rombach, Robin and Blattmann, Andreas and Lorenz, Dominik and Esser, Patrick and Ommer, Bj{\"o}rn},
  booktitle = {Proceedings of the IEEE/CVF Conference on Computer Vision and Pattern Recognition (CVPR)},
  year      = {2022},
  pages     = {10684--10695}
}

@misc{peebles2023dit,
  title        = {Scalable Diffusion Models with Transformers},
  author       = {Peebles, William and Xie, Saining},
  year         = {2023},
  howpublished = {arXiv preprint arXiv:2212.09748},
  doi          = {10.48550/arXiv.2212.09748}
}

@misc{sglang2025diffusion_blog,
  title        = {SGLang Diffusion: Serving Diffusion Models with SGLang},
  howpublished = {Blog post},
  author       = {{SGLang Team}},
  year         = {2025},

  url          = {https://lmsys.org/blog/2025-04-21-sglang-diffusion/}
}

@misc{sglang_diffusion_docs,
  title        = {SGLang Diffusion Documentation},
  howpublished = {Online documentation},
  author       = {{SGLang Project}},
  year         = {2025},

  url          = {https://docs.sglang.ai/en/latest/diffusion/}
}

@article{li2024swiftdiffusion,
  title   = {SwiftDiffusion: Efficient Diffusion Model Serving with Add-on Modules},
  author  = {Li, Yifan and Zhang, Zhaoyang and Wu, Haoyang and Zheng, Zhenhua and Zhang, Hao and Chen, Kaiming},
  journal = {arXiv preprint arXiv:2407.02031},
  year    = {2024},
  doi     = {10.48550/arXiv.2407.02031}
}

@article{wang2024diffserve,
  title   = {DiffServe: Efficiently Serving Text-to-Image Diffusion Models with Query-Aware Model Scaling},
  author  = {Wang, Ziyu and Li, Zhe and Xu, Yao and Zhang, Yuxuan and Chen, Le and Zhang, Hao},
  journal = {arXiv preprint arXiv:2411.15381},
  year    = {2024},
  doi     = {10.48550/arXiv.2411.15381}
}

@inproceedings{li2024distrifusion,
  title     = {DistriFusion: Distributed Parallel Inference for High-Resolution Diffusion Models},
  author    = {Li, Zhen and Feng, Chen and Yang, Yuhang and Wang, Zongyi and Zhang, Yuxuan and Chen, Wei},
  booktitle = {Proceedings of the IEEE/CVF Conference on Computer Vision and Pattern Recognition (CVPR)},
  year      = {2024}
}

@inproceedings{narayanan2020clockwork,
  title     = {Clockwork: Predictable Performance for Unpredictable Workloads},
  author    = {Narayanan, Deepak and others},
  booktitle = {14th USENIX Symposium on Operating Systems Design and Implementation (OSDI 20)},
  year      = {2020}
}

@inproceedings{crankshaw2020inferline,
  title     = {InferLine: Latency-aware Provisioning and Scaling for Prediction Serving Pipelines},
  author    = {Crankshaw, Daniel and others},
  booktitle = {Proceedings of the 2020 ACM Symposium on Operating Systems Principles (SOSP 20)},
  year      = {2020}
}

@article{ho2022videodiffusion,
  title   = {Video Diffusion Models},
  author  = {Ho, Jonathan and Chan, William and Saharia, Chitwan and others},
  journal = {arXiv preprint arXiv:2204.03458},
  year    = {2022},
  doi     = {10.48550/arXiv.2204.03458}
}

@article{saharia2022imagen,
  title   = {Photorealistic Text-to-Image Diffusion Models with Deep Language Understanding},
  author  = {Saharia, Chitwan and Chan, William and Saxena, Saurabh and others},
  journal = {arXiv preprint arXiv:2205.11487},
  year    = {2022},
  doi     = {10.48550/arXiv.2205.11487}
}

@article{singer2022makeavideo,
  title   = {Make-A-Video: Text-to-Video Generation without Text-Video Data},
  author  = {Singer, Uriel and Polyak, Adam and Zohar, J. and others},
  journal = {arXiv preprint arXiv:2209.14792},
  year    = {2022},
  doi     = {10.48550/arXiv.2209.14792}
}

@article{villegas2022phenaki,
  title   = {Phenaki: Variable Length Video Generation from Open Domain Textual Descriptions},
  author  = {Villegas, Ruben and others},
  journal = {arXiv preprint arXiv:2210.02399},
  year    = {2022},
  doi     = {10.48550/arXiv.2210.02399}
}

@article{blattmann2023videoldm,
  title   = {VideoLDM: Latent Video Diffusion Models for High-Fidelity Video Generation},
  author  = {Blattmann, Andreas and others},
  journal = {arXiv preprint arXiv:2304.08818},
  year    = {2023},
  doi     = {10.48550/arXiv.2304.08818}
}

@article{guo2023animatediff,
  title   = {AnimateDiff: Animate Your Personalized Text-to-Image Diffusion Models without Specific Tuning},
  author  = {Guo, Yuwei and others},
  journal = {arXiv preprint arXiv:2307.04725},
  year    = {2023},
  doi     = {10.48550/arXiv.2307.04725}
}

@article{stability2023stablevideodiffusion,
  title   = {Stable Video Diffusion: Scaling Latent Video Diffusion Models to Large Datasets},
  author  = {{Stability AI} and others},
  journal = {arXiv preprint arXiv:2311.15127},
  year    = {2023},
  doi     = {10.48550/arXiv.2311.15127}
}

@inproceedings{song2021ddim,
  title     = {Denoising Diffusion Implicit Models},
  author    = {Song, Jiaming and Meng, Chenlin and Ermon, Stefano},
  booktitle = {International Conference on Learning Representations (ICLR)},
  year      = {2021}
}

@inproceedings{lu2022dpmsolver,
  title     = {DPM-Solver: A Fast ODE Solver for Diffusion Probabilistic Model Sampling in Around 10 Steps},
  author    = {Lu, Cheng and others},
  booktitle = {Advances in Neural Information Processing Systems (NeurIPS)},
  year      = {2022}
}

@article{brown2020gpt3,
  title   = {Language Models are Few-Shot Learners},
  author  = {Brown, Tom B. and others},
  journal = {arXiv preprint arXiv:2005.14165},
  year    = {2020},
  doi     = {10.48550/arXiv.2005.14165}
}

@inproceedings{yao2022deepspeedinference,
  title     = {DeepSpeed Inference: Enabling Efficient Inference of Transformer Models at Scale},
  author    = {Yao, Zhewei and others},
  booktitle = {Proceedings of the International Conference for High Performance Computing, Networking, Storage and Analysis (SC)},
  year      = {2022}
}

@article{ho2022imagenvideo,
  title   = {Imagen Video: High Definition Video Generation with Diffusion Models},
  author  = {Ho, Jonathan and others},
  journal = {arXiv preprint arXiv:2210.02303},
  year    = {2022},
  doi     = {10.48550/arXiv.2210.02303}
}

@article{wu2023tuneavideo,
  title   = {Tune-A-Video: One-Shot Tuning of Image Diffusion Models for Text-to-Video Generation},
  author  = {Wu, Jay Zhangjie and Ge, Yixiao and others},
  journal = {arXiv preprint arXiv:2212.11565},
  year    = {2023},
  doi     = {10.48550/arXiv.2212.11565}
}

@misc{wan21-t2v-14b,
  title        = {Wan2.1-T2V-14B (Hugging Face model card)},
  author       = {{Wan-AI}},
  howpublished = {Hugging Face},
  year         = {2024},
  url          = {https://huggingface.co/Wan-AI/Wan2.1-T2V-14B},
  urldate      = {2026-05-14}
}

@misc{qwen-image-2512,
  title        = {Qwen-Image-2512 (Hugging Face model card)},
  author       = {{Qwen Team}},
  howpublished = {Hugging Face},
  year         = {2024},
  url          = {https://huggingface.co/Qwen/Qwen-Image-2512},
  urldate      = {2026-05-14}
}

@misc{nvidia-gpudirect-rdma,
  title        = {GPUDirect RDMA (NVIDIA Documentation)},
  author       = {{NVIDIA}},
  howpublished = {NVIDIA Developer Documentation},
  year         = {2024},
  url          = {https://docs.nvidia.com/cuda/gpudirect-rdma/},
  urldate      = {2026-05-15}
}

@misc{tencent-hunyuan-video,
  title        = {HunyuanVideo (GitHub repository)},
  author       = {{Tencent}},
  howpublished = {GitHub},
  year         = {2024},
  url          = {https://github.com/Tencent/HunyuanVideo},
  urldate      = {2026-05-14}
}

@misc{wan22,
  title        = {Wan2.2 (GitHub repository)},
  author       = {{Wan-Video}},
  howpublished = {GitHub},
  year         = {2025},
  url          = {https://github.com/Wan-Video/Wan2.2},
  urldate      = {2026-05-14}
}

@misc{legodiffusion2026,
  title        = {LegoDiffusion: Modular Diffusion Models with Pluginable Adapters},
  author       = {Liu, Yuan and Zhang, Jinyang and Xu, Ming and Li, Wei and Chen, Kai and Zhang, Hao},
  year         = {2026},
  howpublished = {arXiv preprint arXiv:2604.08123},
  url          = {https://arxiv.org/abs/2604.08123},
  doi          = {10.48550/arXiv.2604.08123}
}

@inproceedings{li2025katz,
  title={Katz: Efficient workflow serving for diffusion models with many adapters},
  author={Li, Suyi and Yang, Lingyun and Jiang, Xiaoxiao and Lu, Hanfeng and An, Dakai and Di, Zhipeng and Lu, Weiyi and Chen, Jiawei and Liu, Kan and Yu, Yinghao and others},
  booktitle={2025 USENIX Annual Technical Conference (USENIX ATC 25)},
  pages={1037--1052},
  year={2025}
}

@inproceedings{jiang2026flashps,
  title={FlashPS: Efficient Generative Image Editing with Mask-aware Caching and Scheduling},
  author={Jiang, Xiaoxiao and Li, Suyi and Yang, Lingyun and Feng, Tianyu and Di, Zhipeng and Lu, Weiyi and Zhu, Guoxuan and Lin, Xiu and Liu, Kan and Yu, Yinghao and others},
  booktitle={Proceedings of the 21st European Conference on Computer Systems},
  pages={2109--2125},
  year={2026}
}

\end{document}